\documentclass[prd,aps,preprint,amsmath,nofootinbib,amssymb,eqsecnum,showkeys,tightenlines]{revtex4}
\usepackage{amsmath, latexsym, amssymb, graphicx, color, multirow}
\usepackage{gensymb}
\usepackage{graphicx, bm}
\usepackage{subfigure}
\usepackage{mathrsfs}
\usepackage{verbatim}
\usepackage[final]{pdfpages}
\usepackage[sc]{mathpazo}
\usepackage{graphicx}
\usepackage{color}
\usepackage{epstopdf}
\usepackage{subfigure}
\usepackage{hyperref}

\begin{document}

\title{$SU(3)_C\times SU(3)_L\times U(1)_X$ model from $SU(6)$}

\author{Tianjun Li$^{1,2}$}
\email{tli@mail.itp.ac.cn}

\author{Junle Pei$^{1,2}$}
\email{peijunle@mail.itp.ac.cn}

\author{Fangzhou Xu$^{3}$}
\email{xfz14@mails.tsinghua.edu.cn}

\author{Wenxing Zhang$^{1,2}$}
\email{zhangwenxing@mail.itp.ac.cn}

\affiliation{$^1$CAS Key Laboratory of Theoretical Physics, Institute of Theoretical Physics, Chinese Academy of Sciences, Beijing 100190, China}
\affiliation{$^2$ School of Physical Sciences, University of Chinese Academy of Sciences, No.~19A Yuquan Road, Beijing 100049, China}
\affiliation{$^3$  Institute of Modern Physics and Center for High Energy Physics, Tsinghua University, Beijing 100084, China }

\begin{abstract}

We propose the  $SU(3)_C\times SU(3)_L\times U(1)_X$ model arising from $SU(6)$ breaking. 
One family of the Standard Model (SM) fermions arises from two $\bar{6}$  representations
and one $15$ representation of $SU(6)$ gauge symmetry. To break the 
$SU(3)_C\times SU(3)_L\times U(1)_X$ gauge symmetry down to the SM, we introduce 
three $SU(3)_L$ triplet Higgs fields, where two of them come from the $\bar{6}$ representation
while the other one from the $15$ representation. We study the gauge boson masses and Higgs boson mass
in detail, and find that the vacuum expectation value (VEV) of the Higgs field for 
$SU(3)_L\times U(1)_X$ gauge symmetry breaking is around 10 TeV. The neutrino masses and mixing
can be generated via the littlest inverse seesaw mechanism. In particular, we have normal hierarchy
for neutrino masses and the lightest active neutrino is massless. Also, we consider constraints 
from the charged lepton flavor changing decays as well. 
Furthermore, introducing two $SU(3)_L$ adjoint fermions, 
one $SU(3)_C$ adjoint scalar, and one $SU(3)_L$ triplet scalar,
we can achieve gauge coupling unification within 1\%. These extra particles can 
 provide a dark matter candidate as well.
\end{abstract}

\maketitle

\section{Introduction}

The Standard Model (SM) has made a great achievement in explaining the experimental result. 
However, many significant problems remain to be answered. 
Two of the most import issues are the fermion generation and the $U(1)_Y$ hypercharge. 
Since the SM did not explain the origin of the hypercharge, one may expect that 
the quantum number comes from a bigger group, for example, the grand unified theory (GUT). 
In the traditional $SU(3)_C\times SU(3)_L\times U(1)_X$ (331) model, it successfully explained 
why there are three generations by tactfully eliminating $SU(3)_L$ gauge anomalies. 
However, the $U(1)_X$ number is given by hand just like $U(1)_Y$ in the SM, 
which is not satisfying and inspires us to embed the 331 model into a bigger group 
to understand the $U(1)_X$ number more naturally.
In this paper, we shall propose a 331 model generated from a $SU(6)$ model, 
where the $U(1)_X$ charge is determined from the $SU(6)$ breaking. 

In the traditional 331 models~\cite{PhysRevD.22.738, PhysRevD.28.540, Montero:1992jk, Foot:1994ym, Hoang:1995vq, Pleitez:1994pu, Hoang:1997su, Dong:2013ioa, Pisano:1991ee, PhysRevLett.69.2889, Foot:1992rh, Tonasse:1996cx, Nguyen:1998ui, Fonseca:2016tbn, PhysRevD.19.2746, PhysRevD.8.484, PhysRevD.10.1310, Huang:1993qx, Huang:1994zg, Cao:2016uur, Boucenna:2015zwa}, 
the left-handed lepton and one left-handed quark triplet are in the $\bar{3}$  
antifundamental representation of $SU(3)_L$, 
while two left-handed quark triplets are in the $3$ fundamental representation. Thus, we must 
have three generations of leptons in order to cancel gauge anomalies. 
The electric charge operator could be calculated from the diagonal generators of $SU(3)_L \times U(1)_X$ as follows
\begin{equation}\label{charge}
Q=T_3+\beta T_8 + X~.~\,
\end{equation}
Previous models can be classified via the $\beta$ value. For models with 
$\beta=\frac{1}{\sqrt{3}}$~\cite{Pisano:1991ee, Huitu:2019mdr, Huitu:2019kbm, Ponce:2002sg, Dong:2006mg, Boucenna:2015zwa}, 
there are at least three scalars [see the following Eq.~(\ref{higgs})] in Higgs sector in order to 
break $SU(3)_L$ to $U(1)_{EM}$ and generate all the SM fermion and gauge vector masses at tree level. 
In these models, according to Eq.~(\ref{charge}), $Q=\pm diag[\frac{2}{3}+X, -\frac{1}{3}+X, -\frac{1}{3}+X]$ 
(there could be a minus sign for $\bar{3}$ multiplets), all the representations must contain two particles 
with the same charge. For Higgs fields which contain two zero-charged particles, 
there must be two of them in the same representation.  

For models with $\beta=\sqrt{3}$~\cite{PhysRevLett.69.2889, Pisano:1991ee, Foot:1992rh}, it is obvious that 
all the three scalar triplets are all in different representations, because $Q=\pm diag[1+X, X, -1+X]$ for 
particles in (anti)fundamental representation. Moreover, to generate all charged fermion masses in the tree level,
 we need three scalar triplets and one scalar sextet. Such models also contain exotic charged particles 
such as double charged Higgs and quarks with charge $\pm\frac{5}{3}$ and $\pm\frac{4}{3}$. In particular,
there exists the Landau pole problem for $U(1)_X$ not far from the TeV scale.

{We propose the  $SU(3)_C\times SU(3)_L\times U(1)_X$ model, which can be obtained
 from the $SU(6)$ breaking. Such kind of models have been studied previously~\cite{Deppisch:2016jzl,Sen:1983xj}} One family of the SM fermions arises from two $\bar{6}$  representations
and one $15$ representation of $SU(6)$ gauge symmetry. To break the 
$SU(3)_C\times SU(3)_L\times U(1)_X$ gauge symmetry down to the SM gauge symmetry, we introduce 
three $SU(3)_L$ triplet Higgs fields, where two of them arise from $\bar{6}$ representation
while the other one from $15$ representation. We discuss the gauge boson masses and Higgs boson mass in detail,
and show that the vacuum expectation value (VEV) of the Higgs field for 
$SU(3)_L\times U(1)_X$ gauge symmetry breaking is around 10 TeV. We explain the neutrino masses and mixing
 via the littlest inverse seesaw mechanism. Especially, the normal hierarchy for neutrino masses is realized
and the lightest active neutrino is massless. Moreover, we study constraints from the charged lepton
flavor changing decays as well. Furthermore, introducing two $SU(3)_L$ adjoint fermions, 
one $SU(3)_C$ adjoint scalar, and one $SU(3)_L$ triplet scalar,
we can achieve gauge coupling unification within 1\%. These extra particles can 
give us a dark matter candidate as well.

The paper is organized as follows. In Sec.~\ref{sec::model_building}, we present the models and Yukawa terms. 
The gauge sector and Higgs sector are studied in Sec.~\ref{sec::gauge_boson} and Sec.~\ref{sec::higgs_sector},
 respectively.  We discuss the neutrino masses and mixing, as well as the charged lepton flavor changing decays 
 in Sec.~\ref{sec::neutrino}. In Sec.~\ref{sec::gauge_coupling_unification}, we consider
 gauge coupling unification and dark matter candidate.
Our conclusion is in Sec.~\ref{sec::conclusion}.

\section{The $SU(3)_C \times SU(3)_L \times U(1)_X$ Model}\label{sec::model_building}

In our 3-3-1 model, the $SU(3)_C \times SU(3)_L \times U(1)_X$ gauge group arises
 from a large $SU(6)$ gauge group. The $U(1)_X$ charge operator for the $6$ representation 
of the $SU(6)$ group is
\begin{align}
	T_{U(1)_X}=\frac{1}{2\sqrt{3}}{\rm diag}[-1,-1,-1,~1,~1,~1]~.
\end{align}
The following representations of the $SU(6)$ group can be decomposed into representations of
 the $SU(3)_C \times SU(3)_L \times U(1)_X$ group as below
\begin{align}
	6&\to (3,~1,~\frac{-1}{2\sqrt{3}})~\bigoplus~(1,~3,~\frac{1}{2\sqrt{3}})~,\\
	\bar{6}&\to (\bar{3},~1,~\frac{1}{2\sqrt{3}})~\bigoplus~(1,~\bar{3},~\frac{-1}{2\sqrt{3}})~,\\
	15&\to (\bar{3},~1,~\frac{-1}{\sqrt{3}})~\bigoplus~(1,~\bar{3},~\frac{1}{\sqrt{3}})~\bigoplus~(3,~3,~0)~.
\end{align}
One family of the SM fermions and extra fermions in our model is
\begin{align}
	\bar{6}&\to
	(1,~\bar{3},~\frac{-1}{2\sqrt{3}})~\bigoplus~(\bar{3},~1,~\frac{1}{2\sqrt{3}})\\
	&\hookrightarrow f_i=\left(e_{Li},-\nu_{Li},~N_{i}\right)~\bigoplus~ d_{Ri}^c~,\\
	\nonumber\\
	\bar{6}^\prime&\to(1,~\bar{3},~\frac{-1}{2\sqrt{3}}) ~\bigoplus~ (\bar{3},~1,~\frac{1}{2\sqrt{3}})\\
	&\hookrightarrow f_i^\prime=\left(e_{Li}^\prime,-\nu_{Li}^\prime,~N_{i}^\prime\right) ~\bigoplus~ D_{Ri}^c~,\\
	\nonumber\\
	15&\to (3,~3,~0)~\bigoplus~(1,~\bar{3},~\frac{1}{\sqrt{3}})~\bigoplus~(\bar{3},~1,~\frac{-1}{\sqrt{3}})\\
	&\hookrightarrow F_i=\left( u_{Li},~d_{Li},~D_{Li} \right) ~\bigoplus~
	Xf_i^c=\left(\nu_{Ri}^{\prime c},~ e_{Ri}^{\prime c},~ e_{Ri}^{c}\right) ~\bigoplus~
	u_{Ri}^c~.
\end{align}
Besides, we have fermions transforming as singlet under the $SU(3)_C\times SU(3)_L\times U(1)_X$ group, 
which are $N_{si}$ and $N^{\prime}_{si}$.
For all the fermions above, $i=1,2,3$ stands for fermion generation. 

In $SU(6)$ model,  two ${\bar 6}$ antifundamental representations and one $15$ antisymmetric representation 
of the fermions are anomaly free. Thus, our model is anomaly free. To be concrete, we can verify it easily
as well. According to~\cite{Diaz:2004fs, Langacker:1980js}, first, for $U(1)_{X}$, we have
\begin{equation}
	\sum_{\psi i} X_{\psi i}=\sum_{\psi i} X_{\psi i}^3= 0~,
\end{equation}
which makes $U(1)_X$ gauge structure anomaly free. For gauge structure of $SU(3)_L/SU(3)_C$, since the number of fermion multiplets in 3 representation equals to the number of fermion multiplets in $\bar{3}$ representation for every generation, it is also anomaly free.

Our model has 3 scalar multiplets coming from two $\bar{6}$ and one $15$ representations of the $SU(6)$ group, which are
\begin{align}\label{higgs}
	15&\to(1,~\bar{3},~\frac{1}{\sqrt{3}}):
	T_u=\frac{1}{\sqrt{2}}
	\left( 
	\begin{matrix}
		v_u+\rho_1+i\sigma_1\\
		\sqrt{2}\chi_1^+\\
		\sqrt{2}\chi_2^+
	\end{matrix}
	\right)
	,~<T_u>=\frac{1}{\sqrt{2}}
	\left( 
	\begin{matrix}
		v_u\\
		0\\
		0
	\end{matrix}
	\right)~,\\
	\bar{6}&\to(1,~\bar{3},~\frac{-1}{2\sqrt{3}}):
	T_d=\frac{1}{\sqrt{2}}
	\left( 
	\begin{matrix}
		\sqrt{2}\xi_2^-\\
		v_d+\rho_2+i\sigma_2\\
		\rho_3+i\sigma_3
	\end{matrix}
	\right)
	,~<T_d>=\frac{1}{\sqrt{2}}
	\left( 
	\begin{matrix}
		0\\
		v_d\\
		0
	\end{matrix}
	\right)~,\\
	\bar{6}&\to (1,~\bar{3},~\frac{-1}{2\sqrt{3}}):
	T=\frac{1}{\sqrt{2}}
	\left(
	\begin{matrix}
		\sqrt{2}\xi_1^-\\
		\rho_4+i\sigma_4\\
		v_t+\rho_5+i\sigma_5
	\end{matrix}
	\right)
	,~<T>=\frac{1}{\sqrt{2}}
	\left( 
	\begin{matrix}
		0\\
		0\\
		v_t
	\end{matrix}
	\right)~.
\end{align}
We use
\begin{align}
	&\tan\theta= \frac{v_u}{v_d}~,\\
	&k=v_t/\sqrt{v_d^2+v_u^2}~,
\end{align}
to parametrize the 3 VEVs, which break the $SU(3)_L\times U(1)_X$ gauge group down to the $U(1)_{EM}$ gauge group. We write the $U(1)_{EM}$ charge operator as
\begin{equation}
	Q=c_1T_{8L}+c_2T_{3L}+c_3XI~.
\end{equation}
Then the condition, which only neutral states of the scalar multiplets can get VEVs, gives
\begin{equation}
	c_1=\frac{c_2}{\sqrt{3}}=\frac{1}{2}c_3~.
\end{equation}
To make SM particles have the same electric charges as in the SM, we find
\begin{equation}
	c_3=\frac{2}{\sqrt{3}}~,
\end{equation}
leading to
\begin{equation}
	Q=\frac{1}{\sqrt{3}}T_{8L}+T_{3L}+\frac{2}{\sqrt{3}}XI~.
\end{equation}

The Yukawa terms and Majorana mass terms of our model are
\begin{equation}
	\begin{aligned}\label{yukawa}
		-\mathcal{L}_{qua}&=y_{ij}^uF_iu_{Rj}^cT_u+y_{ij}^dF_id_{Rj}^cT_d+y_{ij}^DF_iD_{Rj}^cT+H.c,\\
		-\mathcal{L}_{lep}&=y_{ij}^\nu f_if_jT_u+y_{ij}^e f_i Xf_j^cT_d+y_{ij}^{L^\prime}f_i^\prime Xf_j^cT+y_{ij}^Nf_i\bar{T}N_{sj}+y_{ij}^{N^\prime}f^{\prime}_i\bar{T}N^\prime_{sj}+H.c, \\
		-\mathcal{L}_{neu}^{maj}&=\frac{1}{2}
		\left( N_s~~N^\prime_s\right) 
		\left\lbrace 
		\begin{matrix}
			M_s  & M_{ss^\prime}  \\
			M_{ss^\prime}^T  & M^\prime_s 
		\end{matrix}
		\right\rbrace
		\left( 
		\begin{matrix}
			N_s  \\
			N^\prime_s  
		\end{matrix}
		\right) +H.c,
	\end{aligned}
\end{equation}
where $M_s$, $M_{s}^\prime$ and $M_{ss^\prime}$ are $3\times 3$ matrix. For simplicity, 
we do not include all the gauge invariant terms in Eq~(\ref{yukawa}).

\section{Gauge Bosons}\label{sec::gauge_boson}

We write $W_a(a=1,~2,~\dots,~8)$, which is in the adjoint representation of $SU(3)_L$
 in the form of
\begin{align}
	W_aT_a=\frac{1}{2}
	\left[ 
	\begin{matrix}
		W_3+\frac{1}{\sqrt{3}}W_8 & W_1-iW_2 & W_4-iW_5 \\
		W_1+iW_2 & -W_3+\frac{1}{\sqrt{3}}W_8 & W_6-iW_7 \\
		W_4+iW_5 & W_6+iW_7 & -\frac{2}{\sqrt{3}}W_8
	\end{matrix}
	\right]~.
\end{align}
For the adjoint representation of the $SU(3)_L$ group, the electric charge operator is
\begin{align}
	Q=\frac{1}{\sqrt{3}}T_{8L}+T_{3L}=\frac{1}{3}diag[2,-1,~1]~,
\end{align}
giving
\begin{equation}
	[Q,W_aT_a]
	=\frac{1}{\sqrt{2}}
	\left[ 
	\begin{matrix}
		0 & \frac{W_1-iW_2}{\sqrt{2}} & \frac{W_4-iW_5}{\sqrt{2}} \\
		-\frac{W_1+iW_2}{\sqrt{2}} & 0 & 0 \\
		-\frac{W_4+iW_5}{\sqrt{2}} & 0 & 0
	\end{matrix}
	\right]~.
\end{equation}
We thus define $W^{\pm}\equiv \frac{W_1\mp iW_2}{\sqrt{2}}$, 
$W^{\prime\pm} \equiv \frac{W_4\mp iW_5}{\sqrt{2}}$, $V \equiv \frac{W_6- iW_7}{\sqrt{2}}$, and 
$V^* \equiv \frac{W_6+ iW_7}{\sqrt{2}}$. $W^{\pm}$ and $W^{\prime\pm}$ are charged, while $V$ is neutral. 
{Thus, we do not have the double-charged gauge bosons in our model, 
which is a significant phenomenological difference from traditional 331 models.}

With
\begin{align}
	D_\mu=\partial_\mu-ig_LW_\mu^aT_a-ig_XXB_\mu~,
\end{align}
we get
\begin{align}
	&(D^\mu <T>)^\dagger(D_\mu <T>)+(D^\mu <T_d>)^\dagger(D_\mu <T_d>)+(D^\mu <T_u>)^\dagger(D_\mu <T_u>)\nonumber\\
	=&\left( \frac{g_L}{2}\sqrt{v_u^2+v_d^2}\right) ^2W_\mu^+W^{-\mu}+
	\left( \frac{g_L}{2}\sqrt{v_u^2+v_t^2}\right) ^2W_\mu^{\prime+}W^{\prime-\mu}+
	\left( \frac{g_L}{2}\sqrt{v_d^2+v_t^2}\right) ^2V_\mu V^{*\mu}\nonumber\\
	&+\frac{1}{2}
	\left( B~~W_3~~W_8\right) 
	M^2_{mix}
	\left( 
	\begin{matrix}
		B  \\
		W_3  \\
		W_8
	\end{matrix}
	\right)~, \\
	M^2_{mix}&=
	\left\lbrace 
	\begin{matrix}
		\frac{g_X^2}{12}\left(4v_u^2+v_d^2+v_t^2 \right)  & -\frac{\sqrt{3}g_L g_X}{12}\left(2v_u^2+v_d^2 \right)  & -\frac{g_L g_X}{12}\left(2v_u^2-v_d^2+2v_t^2 \right) \\
		-\frac{\sqrt{3}g_L g_X}{12}\left(2v_u^2+v_d^2 \right) & \frac{g_L^2}{4}\left(v_u^2+v_d^2 \right) & \frac{g_L^2}{4\sqrt{3}}\left( v_u^2-v_d^2\right)  \\
		-\frac{g_L g_X}{12}\left(2v_u^2-v_d^2+2v_t^2 \right) & \frac{g_L^2}{4\sqrt{3}}\left( v_u^2-v_d^2\right)  & \frac{g_L^2}{12}\left(v_u^2+v_d^2+4v_t^2 \right) 
	\end{matrix}
	\right\rbrace.
\end{align}
And we get
\begin{align}
	M_W&=\frac{g_L}{2}\sqrt{v_u^2+v_d^2}~,\\
	M_{W^\prime}&=\frac{g_L}{2}\sqrt{v_u^2+v_t^2}~,\\
	M_{V}&=\frac{g_L}{2}\sqrt{v_d^2+v_t^2}~.
\end{align}
To make $W^\pm$, which is the familiar $W^\pm$ gauge boson in the SM, have the right mass, we have
\begin{align}
	\sqrt{v_u^2+v_d^2}~=~246~{\rm GeV}~.
\end{align}
Also, by diagonalizing $M^2_{mix}$, we get
\begin{align}
	M_A&=0~,\\
	M_Z^2&=m_1^2(1-\sqrt{1-\rho})~,\\
	M^2_{Z^\prime}&=m_1^2(1+\sqrt{1-\rho})~,
\end{align}
with
\begin{align}
	m_1^2&=\frac{1}{6}\left(g_L^2\left(v_u^2+v_d^2+v_t^2 \right)+\frac{g_X^2}{4}(4v_u^2+v_d^2+v_t^2)  \right)~, \\
	\rho&=\frac{3g_L^2\left(g_L^2+g_X^2 \right)\left(v_t^2v_u^2+v_t^2v_d^2+v_u^2v_d^2 \right)  }{\left(g_L^2\left(v_u^2+v_d^2+v_t^2 \right)+\frac{g_X^2}{4}(4v_u^2+v_d^2+v_t^2)  \right)^2}~,
\end{align}
where $A$, $Z$, and $Z^\prime$ are the eigenstates of the mixing of $B$, $W_3$, and $W_8$. $A$ and $Z$ are the photon and the $Z$ gauge boson in the SM, respectively.

We also find
\begin{equation}
	B=\frac{g_L}{\sqrt{g_L^2+g_X^2}}A+\cdots~,
\end{equation}
which means $g_X=\frac{2g_Lg_Y}{\sqrt{3g_L^2-g_Y^2}}.$

With the condition that $|k|\gg 1$, we have
\begin{align}
	M_Z&\approx \frac{g_L}{2\cos\theta_W}\sqrt{v_u^2+v_d^2}~,\\
	M_{Z^\prime}&\approx \frac{g_L}{\sqrt{3-\tan^2 \theta_W}}|v_t|~,
\end{align}
and
\begin{align}
	A&=\sqrt{\cos^2\theta_W-\frac{\sin^2 \theta_W}{3}}B+\sin\theta_W W_3+\frac{\sin\theta_W}{\sqrt{3}} W_8~, \\
	Z&\approx -\sin\theta_W\sqrt{1-\frac{\tan^2\theta_W}{3}}B+\cos\theta_W W_3-\frac{\sin\theta_W\tan\theta_W}{\sqrt{3}}W_8~, \\
	Z^\prime&\approx \frac{\tan\theta_W}{\sqrt{3}} B-\sqrt{1-\frac{\tan^2 \theta_W}{3}}W_8~, 
\end{align}
where $\theta_W$ is the Weinberg angle.

According to ~\cite{Sirunyan:2018nnz} $M_{Z^\prime}$ larger than 4.5~TeV, 
$|v_t|$ needs to be larger than 10~TeV. 

\section{Higgs Sector}\label{sec::higgs_sector}
The most general Higgs potential in our model is
\begin{align}
	V_{\rm Higgs}=& -m_1^2\left|T \right|^2 -m_2^2\left|T_d \right|^2-m_3^2\left|T_u \right|^2 \nonumber \\
	&+l_1\left|T \right|^4+ l_2\left|T_d \right|^4+l_3\left|T_u \right|^4 \nonumber\\
	&+l_{13}\left|T \right|^2 \left|T_u \right|^2+l_{12}\left|T \right|^2 \left|T_d \right|^2+l_{23}\left|T_u \right|^2 \left|T_d \right|^2 \nonumber\\
	&+l^{\prime}_{12}\left|T^\dagger T_d \right|^2 {+l^{\prime}_{13}\left|T^\dagger T_u \right|^2+l^{\prime}_{23}\left|T_u^\dagger T_d \right|^2} \nonumber\\
	&+\left(y_1 T^\dagger T_d \left|T\right|^2 +
	y_{2} T^\dagger T_d \left|T_d\right|^2+
	y_{3} T^\dagger T_d \left|T_u\right|^2+H.c.
	\right)  \nonumber\\
	&+\left(-B T^\dagger T_d+AT_uT_dT+y_{12} T^\dagger T_d T^\dagger T_d {+y_{123} T_u^\dagger T_d T_u T ^\dagger}+H.c. \right)~. 
\end{align}
Since $<\frac{\partial V_{\rm Higgs}}{\partial \rho_i}>=0(i=1,2,\dots,5)$, we get 4 independent relations, which are
\begin{align}
	m_1^2&=\frac{l_{12}v_d^2v_t+2l_1v_t^3+\sqrt{2}Av_dv_u+l_{13}v_tv_u^2}{2v_t}~,\\
	m_2^2&=\frac{l_{12}v_t^2v_d+2l_2v_d^3+\sqrt{2}Av_tv_u+l_{23}v_dv_u^2}{2v_d}~,\\
	m_3^2&=\frac{l_{23}v_d^2v_u+2l_3v_u^3+\sqrt{2}Av_tv_d+l_{13}v_uv_t^2}{2v_u}~,\\
	B&=\frac{y_1v_t^2+y_2v_d^2+y_3v_u^2}{2}~.
\end{align}

\subsection{Mixing of $\xi_{1,2}^\pm,~\chi_{1,2}^\pm$}\label{sec::charged_higgs}

From the Higgs potential $V_{\rm Higgs}$, we get
\begin{align}
	V_{Higgs}\ni
&\left( \chi_1^+~~\xi_2^+~~\chi_2^+~~\xi_1^+\right) 
{M_c^2}
\left( 
\begin{matrix}
\chi_1^- ~~ 
\xi_2^-  ~~
\chi_2^- ~~
\xi_1^-
\end{matrix}
\right)^{\text{T}}~,\\
{M_c^2}=&
\left\lbrace 
\begin{matrix}
-\frac{A v_d v_t}{\sqrt{2}v_u} {+\frac{1}{2}l^\prime_{23}v_d^2}  & -\frac{A v_t}{\sqrt{2}}{+\frac{1}{2}l^\prime_{23}v_d v_u} & {\frac{1}{2} y_{123}v_d v_t} & {\frac{1}{2} y_{123}v_d v_u} \\
-\frac{A v_t}{\sqrt{2}} {+\frac{1}{2}l^\prime_{23}v_d v_u}  & -\frac{A v_u v_t}{\sqrt{2}v_d} {+\frac{1}{2}l^\prime_{23} v_u^2} & {\frac{1}{2} y_{123}v_u v_t} & {\frac{1}{2} y_{123}v_u^2} \\
{\frac{1}{2} y_{123}v_d v_t}  & {\frac{1}{2} y_{123}v_u v_t} & -\frac{A v_t v_d}{\sqrt{2}v_u} {+\frac{1}{2}l^\prime_{13}v_t^2} & -\frac{A v_d}{\sqrt{2}} {+\frac{1}{2}l^\prime_{13}v_t v_u} \\
{\frac{1}{2} y_{123}v_d v_u}  & {\frac{1}{2} y_{123}v_u^2} & -\frac{A v_d}{\sqrt{2}} {+\frac{1}{2}l^\prime_{13}v_t v_u} & -\frac{A v_d v_u}{\sqrt{2}v_t} {+\frac{1}{2}l^\prime_{13}v_u^2} 
\end{matrix}
\right\rbrace~.
\end{align}

Eigenstates from the mixing of $\xi_{1,2}^\pm,~\chi_{1,2}^\pm$ are
\begin{align}
	&\eta_1^\pm=-\frac{v_u}{\sqrt{v_u^2+v_t^2}}\chi_2^\pm+\frac{v_t}{\sqrt{v_u^2+v_t^2}}\xi_1^\pm,~~~~m_{\eta_1}^2=0~,\\
	&\eta_2^\pm=-\frac{v_u}{\sqrt{v_u^2+v_d^2}}\chi_1^\pm+\frac{v_d}{\sqrt{v_u^2+v_d^2}}\xi_2^\pm,~~~~m_{\eta_2}^2=0~,
\end{align}
{and massive eigenstates $\eta_3^\pm$, $\eta_4^\pm$. 
The expressions and masses of $\eta_{3}^\pm$ and $\eta_{4}^\pm$ are not given here because 
they are tedious and easy to get.}
Apparently, $\eta_1^\pm$ and $\eta_2^\pm$ are Goldstone bosons.

\subsection{Mixing of $\sigma_i$\label{sec::pesu_scalar_higgs}}

We have
\begin{align}
	V_{\rm Higgs}\ni\frac{1}{2}\left( \sigma_1~~\sigma_2~~\sigma_5~~\sigma_3~~\sigma_4\right) 
	\left\lbrace 
	\begin{matrix}
		-\frac{Av_dv_t}{\sqrt{2}v_u} & -\frac{Av_t}{\sqrt{2}} & -\frac{Av_d}{\sqrt{2}}  & 0 & 0\\
		-\frac{Av_t}{\sqrt{2}} & -\frac{Av_uv_t}{\sqrt{2}v_d} & -\frac{Av_u}{\sqrt{2}} & 0 & 0 \\
		-\frac{Av_d}{\sqrt{2}} & -\frac{Av_u}{\sqrt{2}} & -\frac{Av_dv_u}{\sqrt{2}v_t} & 0 & 0 \\
		0 & 0 & 0 & -\frac{v_t}{v_d}m_{34}^2 & m_{34}^2\\
		0 & 0 & 0 & m_{34}^2 & -\frac{v_d}{v_t}m_{34}^2 
	\end{matrix}
	\right\rbrace
	\left( 
	\begin{matrix}
		\sigma_1  \\
		\sigma_2   \\
		\sigma_5   \\
		\sigma_3  \\
		\sigma_4
	\end{matrix}
	\right)~,
\end{align}
where
\begin{align*}
	m_{34}^{2}=-\frac{1}{2}l^\prime_{12}v_dv_t+\frac{Av_u}{\sqrt{2}} +y_{12}v_dv_t~.
\end{align*}
The eigenstates are
\begin{align}
	a_1=&\frac{v_u}{\sqrt{v_u^2+v_t^2}}\sigma_1-\frac{v_t}{\sqrt{v_u^2+v_t^2}}\sigma_5~,\\
	a_2=&\frac{v_t^2v_u}{\sqrt{\left(v_u^2+v_t^2 \right) \left( v_t^2v_d^2+v_t^2v_u^2+v_u^2v_d^2\right) }}\sigma_1-\frac{v_d\left(v_u^2+v_t^2 \right) }{\sqrt{\left(v_u^2+v_t^2 \right) \left( v_t^2v_d^2+v_t^2v_u^2+v_u^2v_d^2\right)}}\sigma_2\nonumber\\
	&+\frac{v_u^2v_t}{\sqrt{\left(v_u^2+v_t^2 \right) \left( v_t^2v_d^2+v_t^2v_u^2+v_u^2v_d^2\right)}}\sigma_5~,\\
	a_3=&\frac{v_d}{\sqrt{v_d^2+v_t^2}}\sigma_3+\frac{v_t}{\sqrt{v_d^2+v_t^2}}\sigma_4~,\\
	a_4=&-\frac{v_t}{\sqrt{v_d^2+v_t^2}}\sigma_3+\frac{v_d}{\sqrt{v_d^2+v_t^2}}\sigma_4~,\\
	a_5=&\frac{v_tv_d}{\sqrt{v_t^2v_d^2+v_t^2v_u^2+v_u^2v_d^2}}\sigma_1+\frac{v_tv_u}{\sqrt{v_t^2v_d^2+v_t^2v_u^2+v_u^2v_d^2}}\sigma_2+\frac{v_uv_d}{\sqrt{v_t^2v_d^2+v_t^2v_u^2+v_u^2v_d^2}}\sigma_5~,
\end{align}
and their masses satisfy
\begin{align}
	&m_{a_1}^2=m_{a_2}^2=m_{a_3}^2=0~,\\
	&m_{a_4}^2=-\frac{v_d^2+v_t^2}{v_dv_t}m_{34}^2~, \\
	&m_{a_5}^2=-\frac{A\left( v_t^2v_d^2+v_t^2v_u^2+v_u^2v_d^2\right) }{\sqrt{2}v_dv_uv_t}~.
\end{align}
$a_1$, $a_2$ and $a_3$ are Goldstone bosons.

\subsection{Mixing of $\rho_i$\label{sec::scalar_higgs}}

From the Higgs potential $V_{\rm Higgs}$, we have
\begin{align}
	V_{\rm Higgs}&\ni\frac{1}{2}\rho_i[M_\rho^2]_{i,j} \rho_j~,\\
	M^2_\rho&=\left\lbrace 
	\begin{matrix}
		-\frac{Av_dv_t}{\sqrt{2}v_u}+2l_3v_u^2 & \frac{Av_t}{\sqrt{2}}+l_{23}v_dv_u & y_3v_uv_t  & y_3v_uv_d & \frac{Av_d}{\sqrt{2}}+l_{13}v_uv_t\\
		\frac{Av_t}{\sqrt{2}}+l_{23}v_dv_u & -\frac{Av_uv_t}{\sqrt{2}v_d}+2l_2v_d^2 & y_2v_dv_t  & y_2v_d^2 & \frac{Av_u}{\sqrt{2}}+l_{12}v_dv_t\\
	y_3v_uv_t  & y_2v_dv_t &  -\frac{v_t}{v_d}m_{34}^{\prime 2} & -m_{34}^{\prime 2} & y_1v_t^2\\
		y_3v_uv_d & y_2v_d^2 & -m_{34}^{\prime 2} & -\frac{v_d}{v_t}m_{34}^{\prime 2} & y_1v_dv_t\\
			\frac{Av_d}{\sqrt{2}}+l_{13}v_uv_t  & \frac{Av_u}{\sqrt{2}}+l_{12}v_dv_t & y_1v_t^2 & y_1v_dv_t & -\frac{Av_dv_u}{\sqrt{2}v_t}+2l_1v_t^2\\
	\end{matrix}
	\right\rbrace~,\\
	m_{34}^{\prime 2}&=-\frac{1}{2}l^\prime_{12}v_dv_t+\frac{Av_u}{\sqrt{2}} -y_{12}v_dv_t~.
\end{align}
The lightest eigenstate,
\begin{align}
	&h_1=-\frac{v_d}{\sqrt{v_d^2+v_t^2}}\rho_3+\frac{v_t}{\sqrt{v_d^2+v_t^2}}\rho_4~,
\end{align}
is massless, which is a Goldsten boson.

\begin{figure}[thb]
	\begin{center}
		\includegraphics[width=8.25cm]{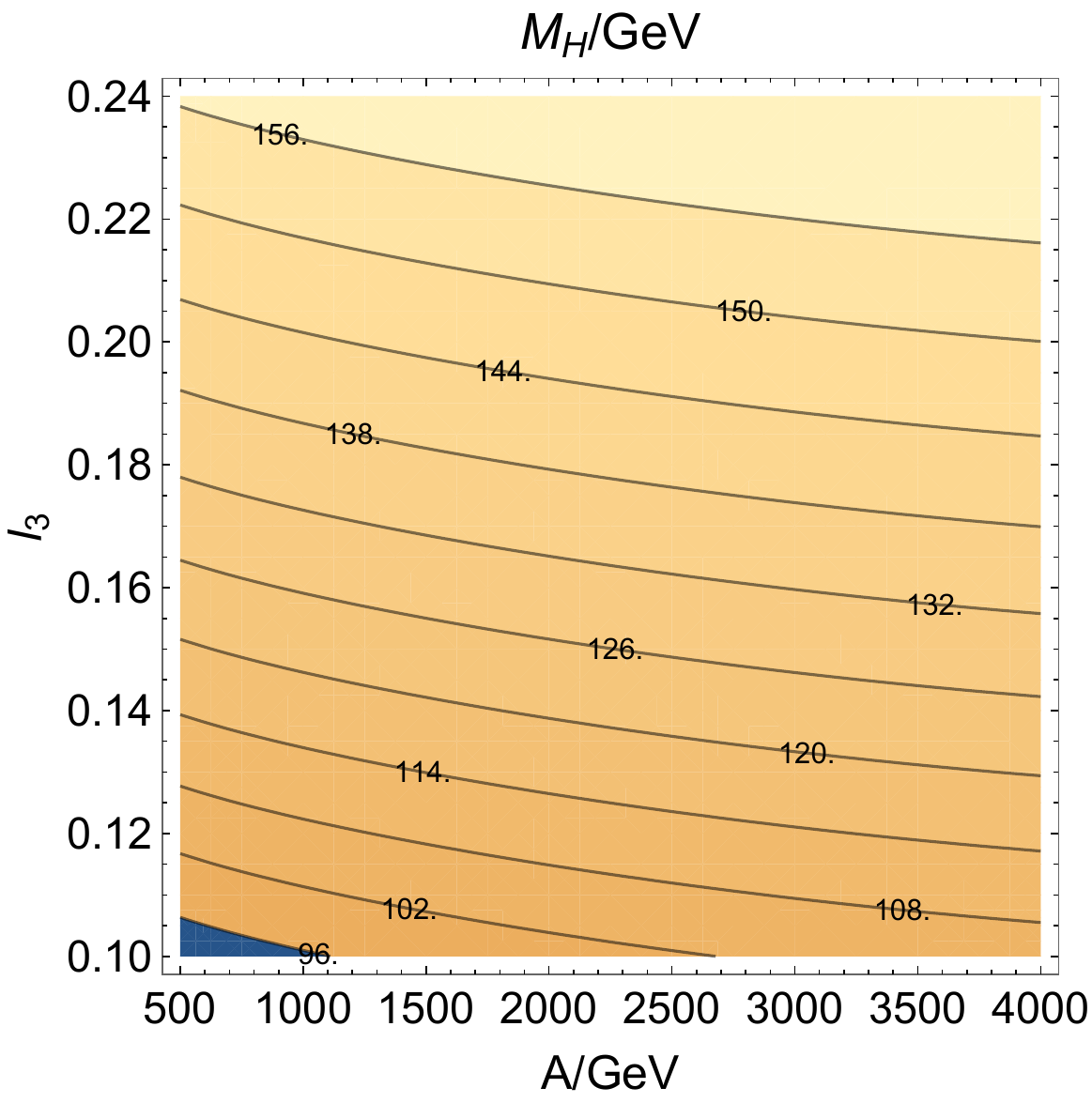}
		\includegraphics[width=8cm]{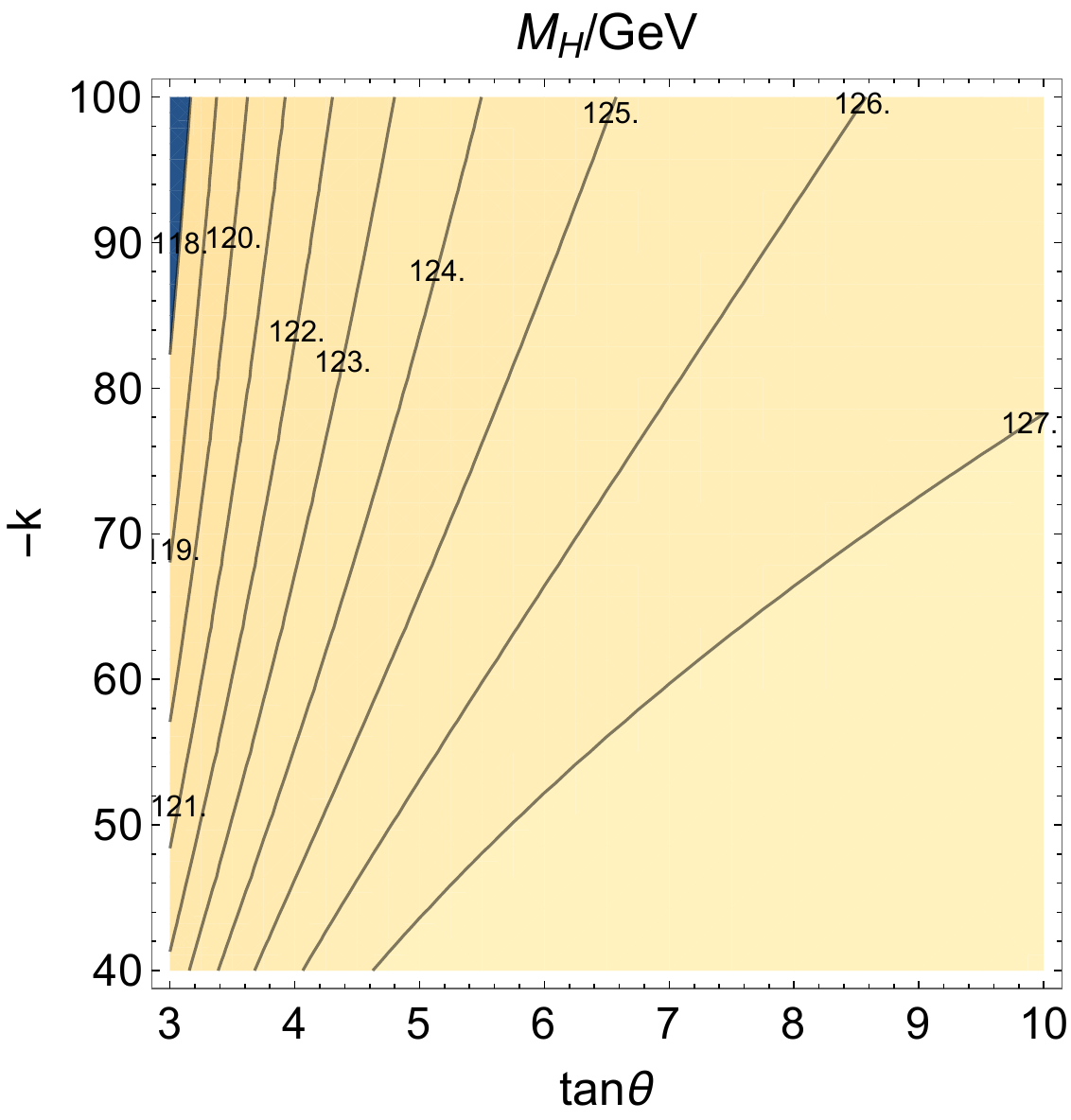}
	\end{center}
	\caption{ Higgs boson mass. On the $A-l_3$ plane, we choose that $\tan\theta=6,~k=-60$, 
and all other dimensionless parameters in $V_{\rm Higgs}$ are $0.1$. On the $\tan\theta-k$ plane, 
we choose that $A=1~{\rm TeV},~l_3=0.16$, and all other dimensionless parameters in $V_{\rm Higgs}$ are $0.1$.}
	\label{higgsm}
\end{figure}

The next to the lightest eigenstate is the SM Higgs boson, whose mass $M_H$ should be 125~GeV. 
The independent parameters in the Higgs potential affecting $M_H$ are $\tan\theta$, $k$, $l_1$, $l_2$, $l_3$, $l_{12}$, $l_{13}$, $l_{23}$, $l_{12}^\prime$, $y_1$, $y_2$, $y_3$, $y_{12}$, and $A$. All these parameters except $A$ are dimensionless. For simplicity, in Fig.~\ref{higgsm}, we show the dependence of $M_H$ on $(A,~l_3)$ and $(\tan\theta,~k)$ respectively while fixing other parameters.  

\section{Neutrino Mass, Mixing, and FCNC}\label{sec::neutrino}
From Eq.~(\ref{yukawa}), the neutrino mass matrix in the basis $(\nu_L,~\nu^\prime_L,~\nu_R^{\prime c},~N,~N_s,~N_s^\prime,~N^\prime)$ is
\begin{align}\label{yuk_matrix}
	M=
	\left[
	\begin{matrix}
		0  & 0  & 0 & \frac{\left({y^{\nu}}^T-y^\nu \right)v_u }{\sqrt{2}} & 0 & 0 & 0 \\
		0  & 0  & \frac{y^{L^\prime} v_t }{\sqrt{2}} & 0 & 0 & 0 & 0  \\
		0  & \frac{{y^{L^\prime }}^T v_t }{\sqrt{2}}  & 0 & \frac{{y^{e }}^T v_d }{\sqrt{2}} & 0 & 0 & 0 \\
		\frac{\left(y^{\nu }-{y^\nu}^T \right)v_u }{\sqrt{2}}  & 0  & \frac{{y^{e }} v_d }{\sqrt{2}} & 0 & \frac{{y^{N }} v_t }{\sqrt{2}} & 0 & 0 \\
		0  & 0  & 0 & \frac{{y^{N }}^T v_t }{\sqrt{2}} & M_s & M_{s s^\prime} & 0  \\
		0  & 0  & 0 & 0 & M_{s s^\prime}^T & M_s^\prime & \frac{{y^{N^\prime }}^T v_t }{\sqrt{2}} \\
		0  & 0  & 0 & 0 & 0 & \frac{{y^{N^\prime }} v_t }{\sqrt{2}} & 0 
	\end{matrix}
	\right].
\end{align}
Every element in $M$ is a $3\times3$ matrix. Because $\left(y^\nu-{y^\nu}^T \right)$ is an antisymmetric matrix, we have
\begin{align}
	&{\rm det}[M]=0~,
\end{align}
which means the lightest neutrino eigenstate is massless.

{For simplicity, we choose $M_{ss'}$ to be a zero matrix.} In the limits of $\tan \theta \gg 1$ and $|k|\gg 1$, we approximately get that $(\nu_L,~N,~N_s)$ are only mixing with themselves and the mass matrix is
\begin{align}M^\prime=
	\left[
	\begin{matrix}
		0 & \frac{\left({y^{\nu}}^T-y^\nu \right)v_u }{\sqrt{2}} & 0  \\
		\frac{\left(y^{\nu }-{y^\nu}^T \right)v_u }{\sqrt{2}}  & 0  & \frac{{y^{N }} v_t }{\sqrt{2}}\\
		0  & \frac{{y^{N }}^T v_t }{\sqrt{2}} & M_s  
	\end{matrix}
	\right].
\end{align}
We define $M_D=\frac{\left({y^{\nu }}^T-{y^\nu}\right)v_u }{\sqrt{2}}$~and~$M_N=\frac{{y^{N }} v_t }{\sqrt{2}}$.
Notice that the situation here looks very similar to the littlest inverse seesaw (LIS) model~\cite{CarcamoHernandez:2019eme, Gavela:2009cd}, in which the elements of $M_s$ are very small to generate the very small neutrino masses. 
Since ${\rm det}[M_D]$ is zero, the lightest  eigenstate of the mixing of $\nu_L$, $N$ and $N_s$  is massless.

The three light eigenvalues of ${M^\prime}^\dagger M^\prime$ forms the SM neutrino mass squares, which are constrained by neutrino oscillation experiments. According to \cite{CarcamoHernandez:2019eme}, in the case that $M_D,M_s \ll M_N$, the three light neutrino mass squares are eigenvalues of $M_\nu^\dagger M_\nu$ with
\begin{equation}
	M_{\nu}=M_D(M_N^T)^{-1}M_s M_N^{-1}M_D^T~.
\end{equation}
For simplicity, we set $M_N$ and $M_s$ to be diagonal, which are
\begin{align}
	M_N=v_t~diag[c_N,~c_N,~c_N]~,
\end{align}
\begin{align}
	M_s=diag[k_1,~k_2,~k_3]~,
\end{align}
Since $M_D$ is antisymmetric, it can be written as
\begin{align}
	M_D=v_u
	\left[
	\begin{matrix}
		0 & d_1 & d_2  \\
		-d_1  & 0  & d_3\\
		-d_2  & -d_3 & 0  
	\end{matrix}
	\right].
\end{align}
So we have
\begin{align}
	M_\nu=\frac{v_u^2}{v_t^2 c_N^2 }
	\left[
	\begin{matrix}
		d_1^2 k_2+d_2^2 k_3  & d_2 d_3 k_3   & -d_1 d_3 k_2\\
		d_2 d_3 d_3 & d_1^2 d_1+d_3^2 k_3 & d_1 d_2 k_1  \\
		-d_1 d_3 k_2  & d_1 d_2 k_1 & d_1^2 k_1+d_3^2 k_3  
	\end{matrix}
	\right].
\end{align}
Suppose eigenvalues of $M_\nu^\dagger M_\nu$ are $ m_{1}^2=0$, $m_{2}^2$, and $m_{3}^2$.
However, we can always rescale $d_i(i=1,2,3)$ and $k_j(j=1,2,3)$ to $10^{-R_D}d_i$ and $R_s k_j$ without changing the neutrino mixing pattern and $\frac{m_3}{m_2}$. But the masses will be changed to $10^{-2R_D}R_s m_i(i=,1,2,3)$.

Because the lightest neutrino in our model is massless, we should choose appropriate values of $a_i$, $k_j$, $c_N$, $\tan\theta$ and $k$ to give
\begin{align}
	U^\dagger_\nu M_{\nu}^\dagger M_{\nu}  U_\nu=diag[	0,~m_2^2=\Delta m_{21}^2,~m_3^2=\Delta m_{31}^2]~,
\end{align}
where $U_\nu$ is parametrized by $\theta_{12},~\theta_{13},~\theta_{23}$, and $\delta$, {\it i.e.}, 
the normal hierarchy (NH) for neutrino masses. We choose 
\begin{align}
	(d_1,~d_2,~d_3)&=10^{-R_D}(0.49,~0.29,~0.82)~,\\
	(k_1,~k_2,~k_3)&=R_s(0.33,~0.038e^{0.36 \pi i},-0.027 e^{0.13 \pi i})~,
\end{align}
where $R_s$ is determined by $\tan\theta$, $k$, $c_N$, and $R_D$ to give the right neutrino masses. For example, when $\tan\theta=6$, $k=-60$, $c_N=-1$ and $R_D=1$, $R_s$ needs to be $1.8\times 10^{-4}GeV$, giving us the three mixing angles, $CP$ violating phase $\delta$ and neutrino masses in Table~\ref{neutrinom}.
\begin{table}[htbp]
	\centering
	$\begin{array}{|c|c|c|c|}
	\hline 
	\text { Observable } & {\text { Model }} & {\text { bpf } \pm 1 \sigma} & {\text { bpf } \pm 1 \sigma} \\ 
	\hline 
	\Delta m^2_{21}(10^{-5} \mathrm{eV}^{2}) & {7.36} & {7.55_{-0.16}^{+0.20}} & {7.39_{-0.20}^{+0.21}}  \\ 
	\hline 
	\Delta m^2_{31}(10^{-3} \mathrm{eV}^{2}) & {2.53} & {2.50\pm 0.03} & {2.525_{-0.031}^{+0.033}}  \\ 
	\hline 
	\theta_{12}^{(l)}(\degree) & {33.83} & {34.5_{-1.0}^{+1.2}} & {33.82_{-0.76}^{+0.78}} \\ 
	\hline 
	\theta_{13}^{(l)}(\degree) & {8.57} & {8.45_{-0.14}^{+0.16}} & {8.61_{-0.13}^{+0.12}}  \\ 
	\hline 
	\theta_{23}^{(l)}(\degree) & {49.82} & {47.9_{-1.7}^{+1.0}} & {49.7_{-1.1}^{+0.9}}\\ 
	\hline 
	\delta_{CP}^{(l)}(\degree) & {-142.05} & {-142_{-27}^{+38}} & {217_{-28}^{+40}}\\ 
	\hline
	\end{array}$
	\caption{ {Model and experimental values of the light active neutrino masses, leptonic mixing angles, and $CP$ violating phase for the scenario of the NH neutrino masses~\cite{deSalas:2017kay, Esteban:2018azc}.  
	} }
    \label{neutrinom}
\end{table}

\begin{figure}[thb]
	\begin{center}
		\includegraphics[width=8.2cm]{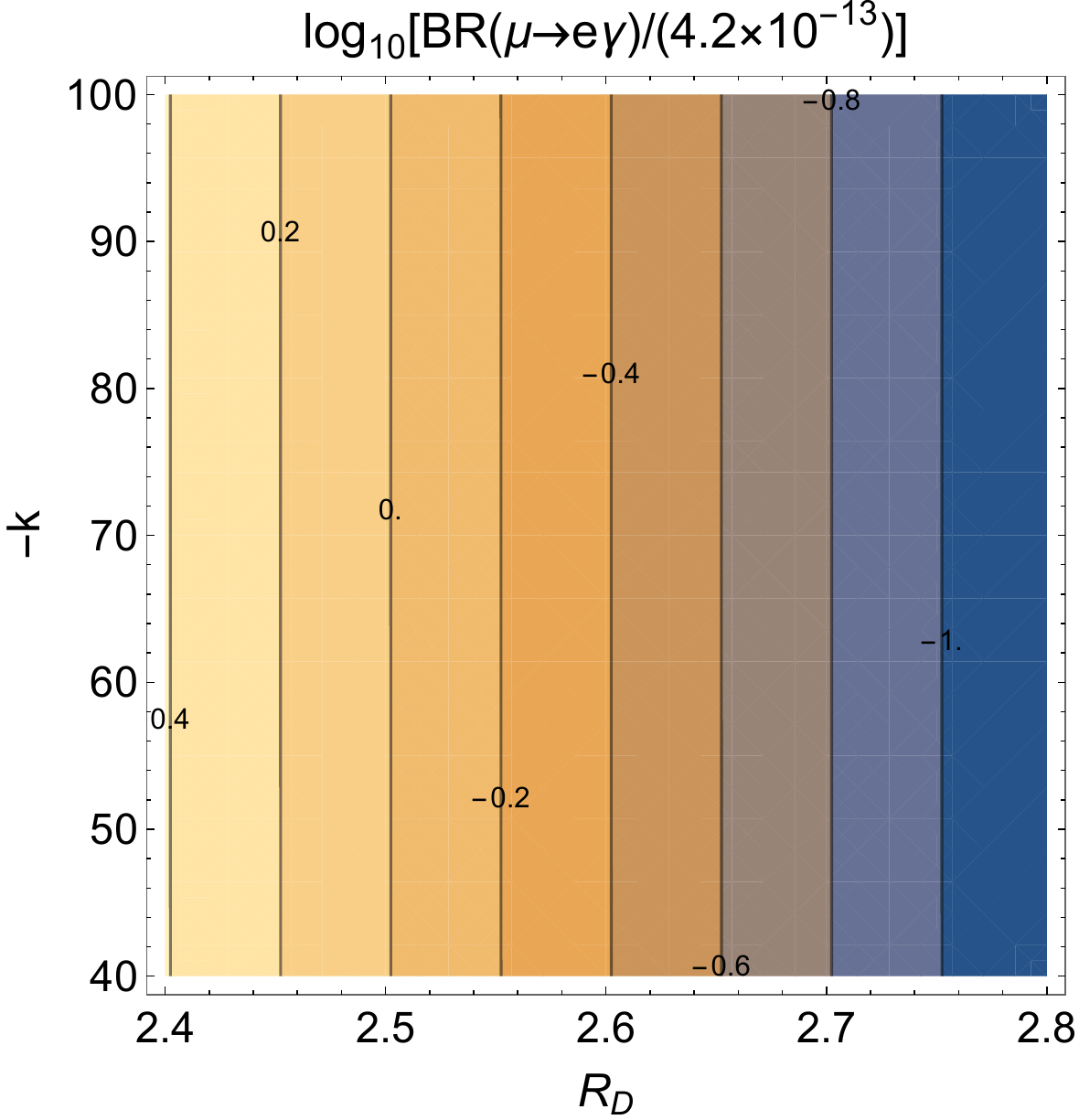}
		\includegraphics[width=8cm]{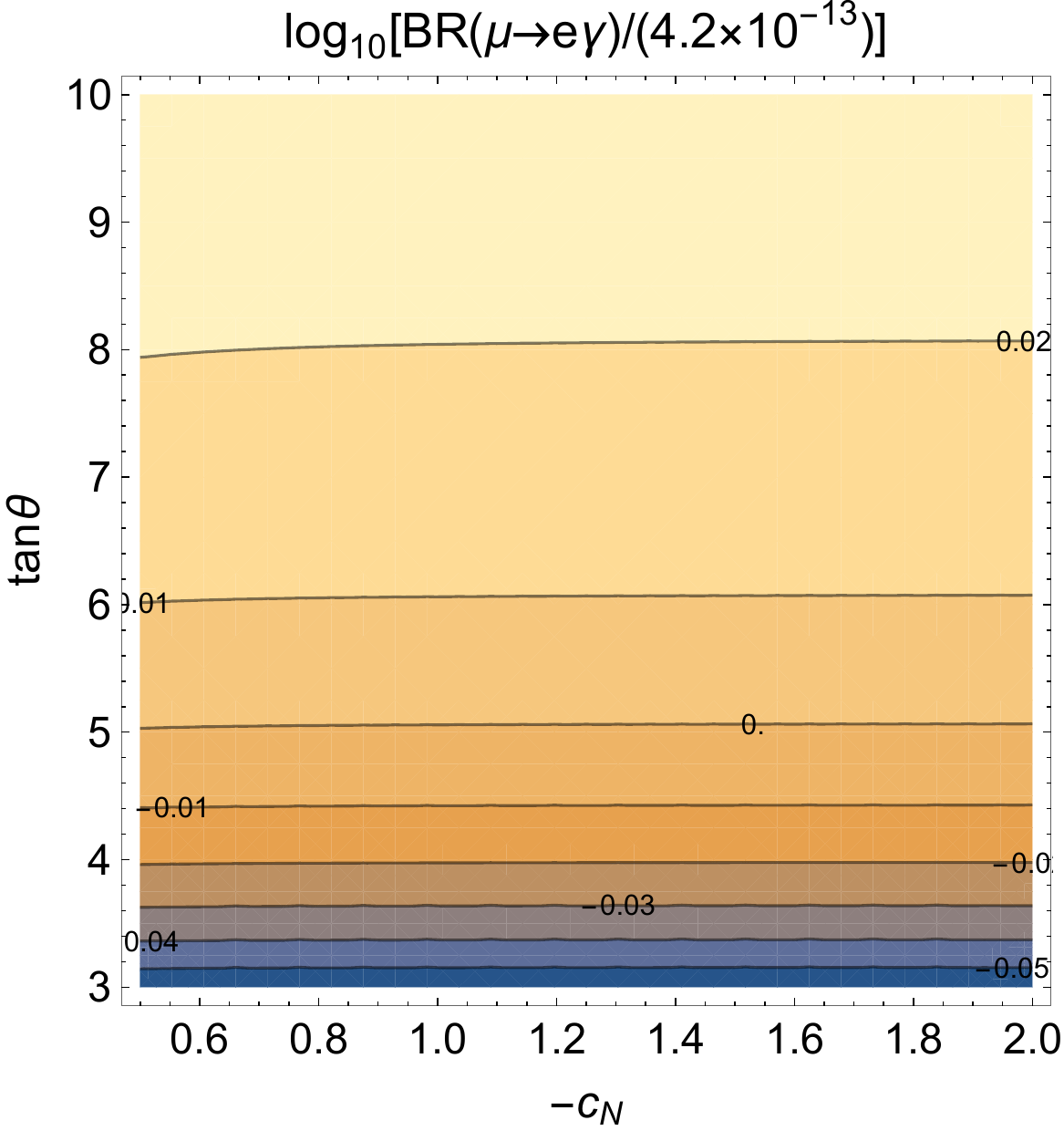}\\
		\includegraphics[width=8cm]{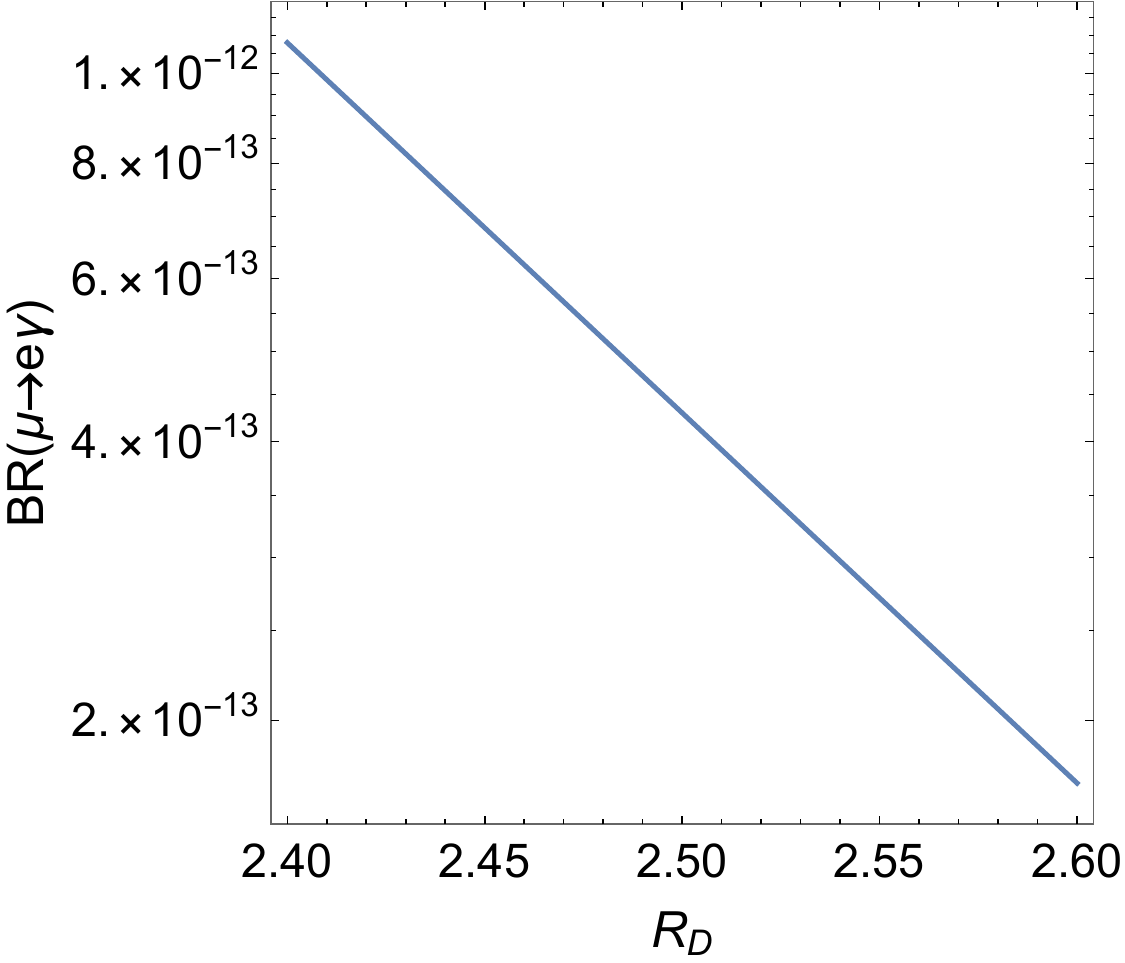}
	\end{center}
	\caption{ $BR(\mu \to e \gamma)$. On the $R_D-k$ plane, we choose that $c_N=-1,~tan\theta=6$. On the $c_N-\tan\theta$ plane, we choose that $k=-60,~R_D=2.5$. The curve of $BR(\mu \to e \gamma)$ is got when $\tan\theta=6,~ k=-60,~c_N=-1$.}
	\label{fcncm}
\end{figure}

Next, we shall discuss the implication of the 3-3-1 model in the charged lepton flavor changing decays. There are in total three processes, which are $\mu \to e \gamma$, $\tau \to \mu \gamma$ and $\tau \to e \gamma$. The branch ratio of lepton $e_i$ decaying to lepton $e_j$ is
\begin{equation}
	BR(e_i \to e_j) = \frac{\alpha_W^3 m_{e_i}^5 s_W^2}{256\pi^2 \Gamma_i} \left|  \sum_{k=1}^{k=9} \left( U^\dagger_{j,k}U_{k,i} G(\frac{m_{N_k}^2}{M_W^2})\frac{1}{M_W^2} + U^\dagger_{j+3,k}U_{k,i+3} \sum_{k=1}^{k=9} G(\frac{m_{N_k}^2}{M_{W'}^2})\frac{1}{M_{W^\prime}^2} \right) \right|^2 ,
\end{equation}
where $U^\dagger{M^\prime}^\dagger M^\prime U=diag[m^2_{N_1},~m^2_{N_2},~\dots,~m^2_{N_9}]$.
Experimental results ask us that the branch ratio of charged lepton decay should satisfy
\begin{align}
	BR(\mu \to e \gamma)\leq 4.2\times 10^{-13}~,\\
	BR(\tau \to \mu \gamma)\leq 4.4\times 10^{-8}~,\\
	BR(\tau \to e \gamma)\leq 3.3\times 10^{-8}~.
\end{align}
Independent parameters  influencing  $BR(e_i \to e_j\gamma)$ are $\tan\theta$, $k$, $R_D$, and $c_N$, while $R_s$ is determined by other parameters to give the right neutrino masses. In Fig.~\ref{fcncm}, we show the dependence of $BR(\mu \to e \gamma)$ on these parameters. We find that $BR(\mu \to e \gamma)$ mainly depends on $R_D$. To make that $BR(\mu \to e \gamma)\leq 4.2\times 10^{-13}$, $R_D$ needs to be larger than $2.5$, which means that $(d_1,~d_2,~d_3)<(1.55,~0.92,~2.81)\times 10^{-3}$. In the case that $R_D\sim 2.5$, $BR(\tau \to \mu \gamma)$, and $BR(\tau \to e \gamma)$ are around $10^{-14}$ and $10^{-13}$ respectively. { Our model has more parameters in the neutrino mass matrix M 
 than traditional 331 models~\cite{PhysRevD.22.738, PhysRevD.28.540, Montero:1992jk, Foot:1994ym, Hoang:1995vq, Pleitez:1994pu, Hoang:1997su, Dong:2013ioa, Pisano:1991ee, PhysRevLett.69.2889, Foot:1992rh, Tonasse:1996cx, Nguyen:1998ui, Fonseca:2016tbn, PhysRevD.19.2746, PhysRevD.8.484, PhysRevD.10.1310, Huang:1993qx, Huang:1994zg, Cao:2016uur, Boucenna:2015zwa}, 
and then it is easier to satisfy these constraints from lepton flavor changing decays.}

\section{Unification of gauge couplings}\label{sec::gauge_coupling_unification}

The renormalization group equation (RGE) for gauge coupling is
\begin{equation}
	\mu\frac{dg_i}{d\mu}=\sum_{n}\frac{1}{(16\pi^2)^n}\beta_i^{(n)},
\end{equation}
where i stands for the i-loop correction in RGE running. In this section, we consider two-loop correction.
Equations of 1-loop and 2-loop corrections are
\begin{align}
	\beta_g^{(1)}&=b_i g_i^3~,\\
	\beta_g^{(2)}
	&=B_{ij}g_j^2+\sum_{\alpha}d_i^{\alpha}Tr[y^{\alpha}y^{\alpha \dagger}]~,
\end{align}
where $\alpha=d,~u,~D,~\nu,~e,~L^\prime,~N,~N^\prime$.

In our model, we get
\begin{align}
	b=&[\frac{13}{2},~-\frac{9}{2},~-5]]~,~~\\
	B=&\left[ 
	\begin{matrix}
		6 & 20 & 12  \\
		\frac{5}{2} & 23  & 12\\
		\frac{3}{2}  & 12 & 12
	\end{matrix}
	\right]~,~~\\
	d=&\left[ 
	\begin{matrix}
		-\frac{3}{4} & -3 & -\frac{3}{4}  & -2 & -\frac{5}{2} & -\frac{5}{2} & -\frac{1}{4} & -\frac{1}{4} \\
		-\frac{3}{2} & -\frac{3}{2}  & -\frac{3}{2} & -4 & -2 & -2 & -\frac{1}{2} & -\frac{1}{2}\\
		-3  & -3 & -3 & 0 & 0 & 0 & 0 & 0
	\end{matrix}
	\right]]~.~
\end{align}

To make gauge couplings unify at the GUT scale, we add two fermion multiplets, $FA$ and $FA^\prime$, as well as two scalar multiplets, $SA$ and  $T^\prime$ in high scale. The details are
\begin{align}
&(1,~8,~0):FA
,~~~~~\Delta b=(0,~2,~0)
,~~~~\Delta B=\left[
\begin{matrix}
0 & 0 & 0  \\
0 & 48  & 0\\
0  & 0 & 0
\end{matrix}
\right]~,~~\\
&(1,~8,~0):FA^\prime
,~~~~~\Delta b=(0,~2,~0)
,~~~~\Delta B=\left[ 
\begin{matrix}
0 & 0 & 0  \\
0 & 48  & 0\\
0  & 0 & 0
\end{matrix}
\right]~,~~\\
&(8,~1,~0):SA
,~~~~~\Delta b=(0,~0,~1)
,~~~~\Delta B=\left[ 
\begin{matrix}
0 & 0 & 0  \\
0 & 0  & 0\\
0  & 0 & 42
\end{matrix}
\right]~,~~\\
&(1,~\bar{3},~\frac{-1}{2\sqrt{3}}):T^\prime
,~~~~~\Delta b=(\frac{1}{12},~\frac{1}{6},~0)
,~~~~\Delta B=\left[ 
\begin{matrix}
\frac{1}{12} & \frac{4}{3} & 0  \\
\frac{1}{6} & \frac{11}{3}  & 0\\
0  & 0 & 0
\end{matrix}
\right]~.~
\end{align}

\begin{figure}[thb]
	\begin{center}
		\includegraphics[width=8cm]{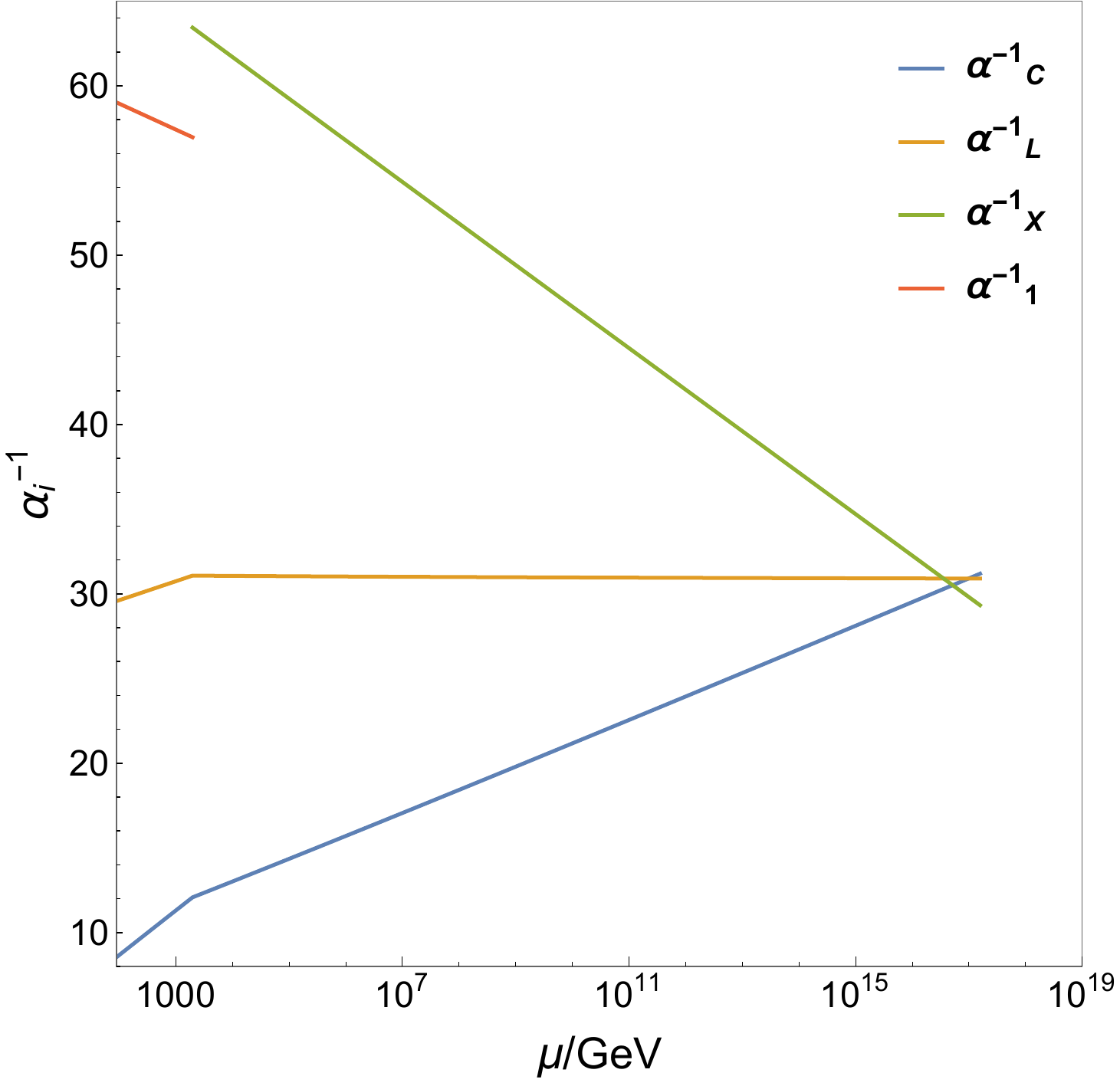}
		\includegraphics[width=7.95cm]{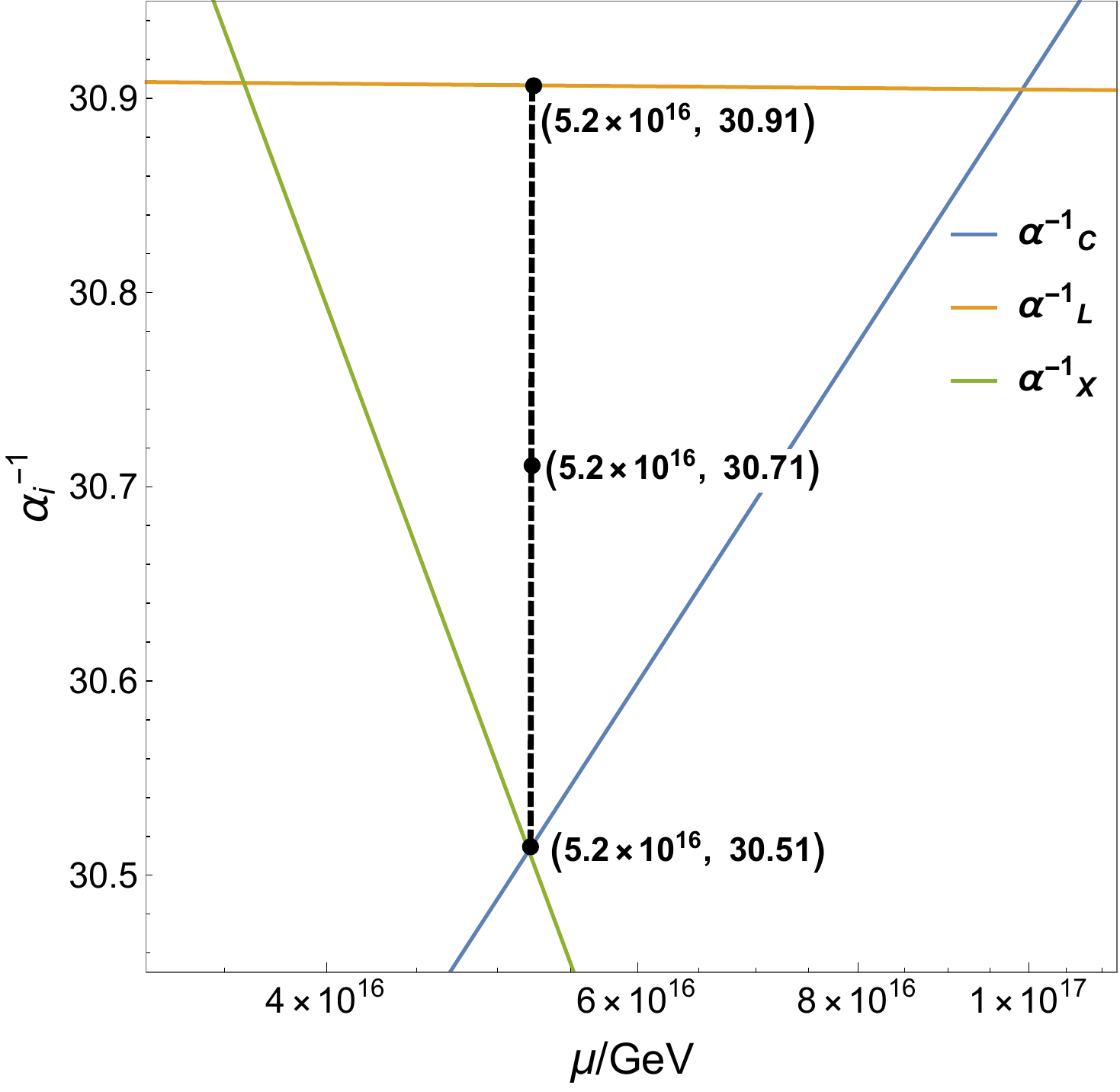}	
	\end{center}
	\caption{ Gauge coupling unification, where $\alpha_i=\frac{g_i^2}{4\pi}$ and $g_1^2=\frac{5}{3}g_Y^2$.}
	\label{uni1}
\end{figure}

\begin{figure}[thb]
	\begin{center}
		\includegraphics[width=10cm]{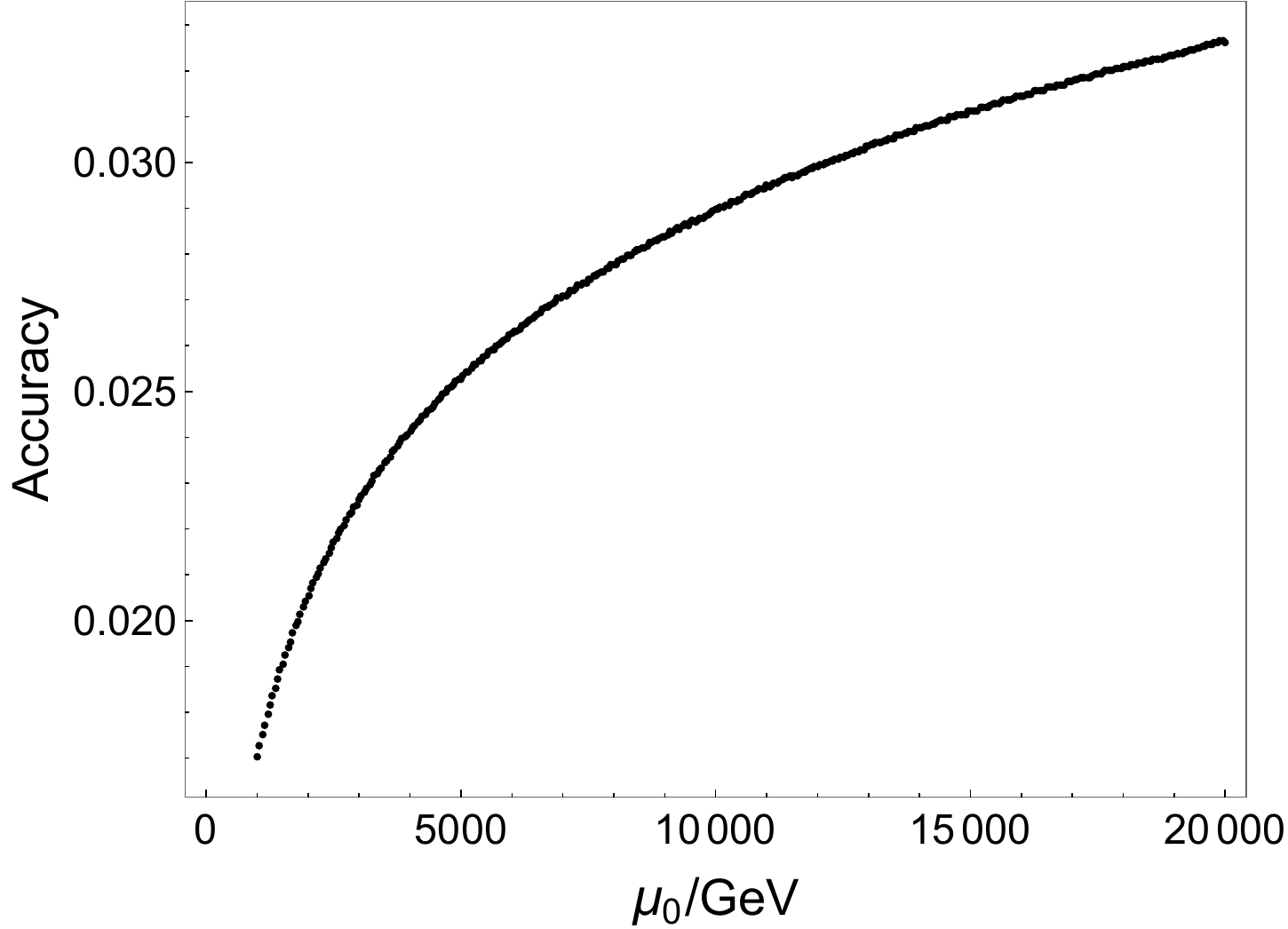}		
	\end{center}
	\caption{{Accuracy of gauge coupling unification. 
We assume that all the new particles beyond the SM have universal mass around the energy scale of $\mu_0$.}}
	\label{uni3}
\end{figure}

\begin{figure}[thb]
	\begin{center}
		\includegraphics[width=8cm]{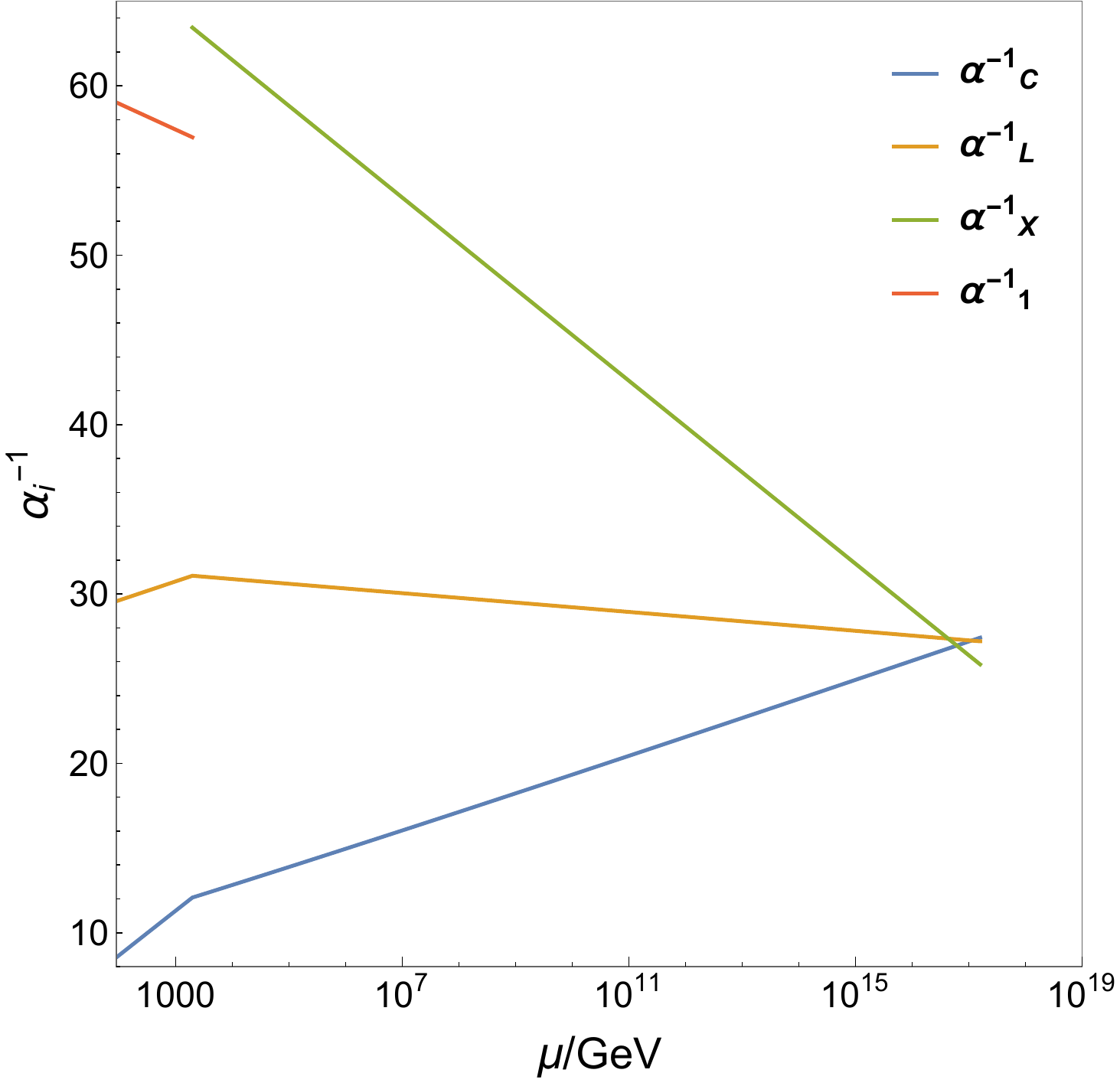}
		\includegraphics[width=8.2cm]{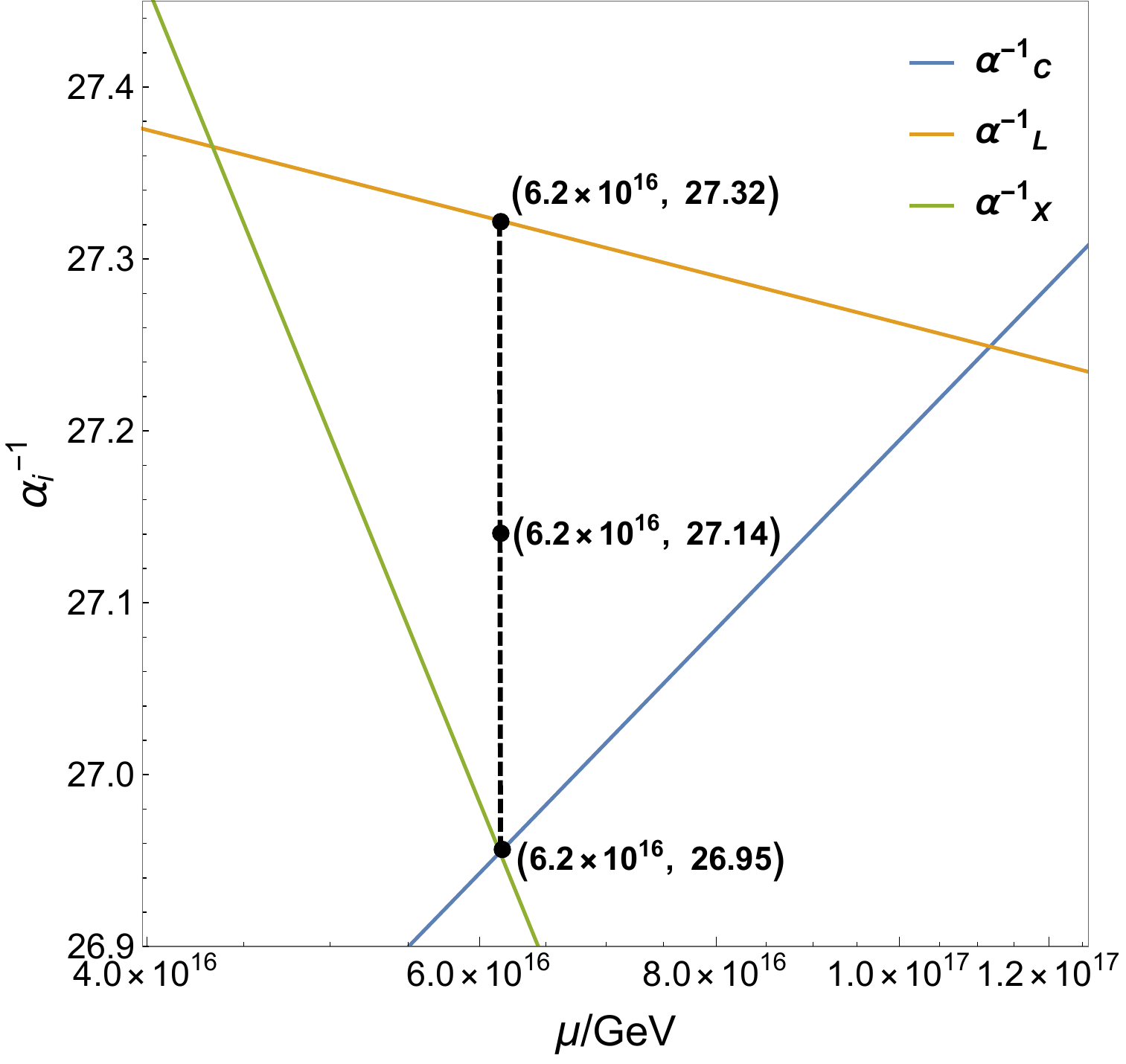}		
	\end{center}
	\caption{ Gauge coupling unification, where $\alpha_i=\frac{g_i^2}{4\pi}$ and $g_1^2=\frac{5}{3}g_Y^2$.}
	\label{uni2}
\end{figure}

$FA$ and $FA'$ can decay via the Yukawa coupling terms $FA f_i (T')^*$, $FA' f_i (T')^*$, 
$FA f'_i (T')^*$, and $FA' f'_i (T')^*$. In principle, we can introduce the $Z_2$ symmetry where
$FA$, $FA'$, and $(T')^*$ are odd while all the other particles are even. Thus, the lightest
particle of $FA$, $FA'$, and $(T')^*$ can be a dark matter candidate. In addition, $SA$
can decay into the SM quarks only at nonrenormalizable level, for example, $SA F_iu_{Rj}^cT_u/M_*$,
$SA F_id_{Rj}^cT_d/M_* $, and $SA F_iD_{Rj}^cT /M_* $. Thus, we have two cases.
First, $SA$ can be a dark matter candidate if $Z_2$ symmetry is imposed to forbid $SA$ decaying to quarks. 
We will leave this part of work in the future. For simplicity, we make all the particles beyond the SM take part in the RGE running at the energy scale of 2~TeV,
then the gauge coupling unification can be satisfied with accuracy of $0.65\%$ at the energy scale of $5.2\times 10^{16}GeV$, which is shown in Fig.~\ref{uni1}. {We define the accuracy of gauge coupling unification as 
$|\alpha^{-1}_X(\mu^\prime)-\alpha^{-1}_C(\mu^\prime)|/\alpha^{-1}_C(\mu^\prime)$ with $\mu^\prime$ satisfying $\alpha^{-1}_X(\mu^\prime)=\alpha^{-1}_L(\mu^\prime)$, which is different from our choice of the accuracy of unification of gauge couplings in Fig.~\ref{uni1} and Fig.~\ref{uni2}. Assuming all the new particles beyond the SM have universal mass around the energy scale of $\mu_0$, we present the relation of the accuracy of gauge coupling unification and $\mu_0$ in Fig.~\ref{uni3}. $\mu_0$ needs to be smaller than 12 TeV to achieve the gauge coupling unification with an accuracy better than $3\%$, which implies that the mass of the dark matter candidate needs to be smaller than 12 TeV. }

Alternatively, to make $SA$ decay, we can add two fermion multiplets in $6$ and $\bar{6}$ representation of the $SU(6)$ gauge group respectively, then the gauge coupling unification can be satisfied with accuracy of $0.68\%$ at the energy scale of $6.2\times 10^{16}GeV$, which is shown in Fig.~\ref{uni2}. Also, we make all the particles beyond 
the SM take part in the RGE running at the energy scale of 2~TeV.

\section{Conclusions}\label{sec::conclusion}

We have proposed a new $SU(3)_C \times SU(3)_L \times U(1)_X$  model, 
in which gauge symmetry can be realized from $SU(6)$ breaking. 
The SM fermions in each of the three generations come from two $\bar{6}$ representations 
and one $15$ representation of the $SU(6)$ gauge group besides two singlets 
of the $SU(3)_C \times SU(3)_L \times U(1)_X$ gauge group. There are three scalar multiplets, where
 two come from $\bar{6}$ representations of $SU(6)$ and one from $15$ representation. 
And their VEVs are $v_u$, $v_d$, and $v_t$, respectively. There are additional gauge bosons, 
${W^\pm}^\prime$, $Z^\prime$, and $V/V^*$, in our model besides the SM gauge bosons. 
$v_t$ needs to be larger than 10~TeV to make the mass of $Z^\prime$ larger than 4.5~TeV. 
It is easy to give the 125~GeV Higgs boson mass when we set all the dimensionless parameters 
in the Higgs potential to be $\sim 0.1$ and $A$ to be $\sim 1~{\rm TeV}$. When $M_{ss^\prime}$ are set 
to be a zero matrix and in the limits of $\tan \theta \gg 1$ and $|k|\gg 1$, the mixing of $\nu_L$, $N$ 
and $N_s$ is the same as in the littlest inverse seesaw model. The lightest neutrino 
in our model is massless. With parameters in $y^\nu$, $y^N$ and $M_s$ set to be appropriate values, 
we obtained the light active neutrino masses, leptonic mixing angles, and $CP$ violating phase 
highly consistent with the experimental data for the scenario of NH neutrino mass. To make 
$BR(\mu \to e \gamma)\leq 4.2\times 10^{-13}$, parameters in $y^\nu$ needs to be smaller 
than $\sim 10^{-3}$, and in this case $BR(\tau \to \mu \gamma)$ and $BR(\tau \to e \gamma)$ 
are around $10^{-14}$ and $10^{-13}$, respectively. With additional two fermion multiplets, $FA$ and $FA^\prime$, 
as well as two scalar multiplets, $SA$ and  $T^\prime$, the gauge coupling unification can be realized
 with accuracy of $0.65\%$ at the energy scale of $5.2\times 10^{16}~{\rm GeV}$. 
$SA$ can be a dark matter candidate if $Z_2$ symmetry is imposed. Alternatively, 
we can add two fermionic multiplets in $6$ and $\bar{6}$ representations of the $SU(6)$ gauge group 
to make $SA$ decay, then the gauge coupling unification can be satisfied with accuracy of $0.68\%$ 
at the energy scale of $6.2\times 10^{16}~{\rm GeV}$.

\begin{acknowledgments}

We would like to thank Bhaskar Dutta and Ilia Gogoladze for helpful discussions in 
the early stage of this project.
This research was supported  by the Projects No.~11847612 and No.~11875062 supported by the 
National Natural Science Foundation of China, and the Key Research Program of Frontier Science, CAS.

\end{acknowledgments}

\bibliography{refer}

\begin{thebibliography}{34}
\expandafter\ifx\csname natexlab\endcsname\relax\def\natexlab#1{#1}\fi
\expandafter\ifx\csname bibnamefont\endcsname\relax
  \def\bibnamefont#1{#1}\fi
\expandafter\ifx\csname bibfnamefont\endcsname\relax
  \def\bibfnamefont#1{#1}\fi
\expandafter\ifx\csname citenamefont\endcsname\relax
  \def\citenamefont#1{#1}\fi
\expandafter\ifx\csname url\endcsname\relax
  \def\url#1{\texttt{#1}}\fi
\expandafter\ifx\csname urlprefix\endcsname\relax\def\urlprefix{URL }\fi
\providecommand{\bibinfo}[2]{#2}
\providecommand{\eprint}[2][]{\url{#2}}

\bibitem[{\citenamefont{Singer et~al.}(1980)\citenamefont{Singer, Valle, and
  Schechter}}]{PhysRevD.22.738}
\bibinfo{author}{\bibfnamefont{M.}~\bibnamefont{Singer}},
  \bibinfo{author}{\bibfnamefont{J.~W.~F.} \bibnamefont{Valle}},
  \bibnamefont{and}
  \bibinfo{author}{\bibfnamefont{J.}~\bibnamefont{Schechter}},
  \bibinfo{journal}{Phys. Rev. D} \textbf{\bibinfo{volume}{22}},
  \bibinfo{pages}{738} (\bibinfo{year}{1980}).

\bibitem[{\citenamefont{Valle and Singer}(1983)}]{PhysRevD.28.540}
\bibinfo{author}{\bibfnamefont{J.~W.~F.} \bibnamefont{Valle}} \bibnamefont{and}
  \bibinfo{author}{\bibfnamefont{M.}~\bibnamefont{Singer}},
  \bibinfo{journal}{Phys. Rev. D} \textbf{\bibinfo{volume}{28}},
  \bibinfo{pages}{540} (\bibinfo{year}{1983}).

\bibitem[{\citenamefont{Montero et~al.}(1993)\citenamefont{Montero, Pisano, and
  Pleitez}}]{Montero:1992jk}
\bibinfo{author}{\bibfnamefont{J.~C.} \bibnamefont{Montero}},
  \bibinfo{author}{\bibfnamefont{F.}~\bibnamefont{Pisano}}, \bibnamefont{and}
  \bibinfo{author}{\bibfnamefont{V.}~\bibnamefont{Pleitez}},
  \bibinfo{journal}{Phys. Rev.} \textbf{\bibinfo{volume}{D47}},
  \bibinfo{pages}{2918} (\bibinfo{year}{1993}), \eprint{hep-ph/9212271}.

\bibitem[{\citenamefont{Foot et~al.}(1994)\citenamefont{Foot, Long, and
  Tran}}]{Foot:1994ym}
\bibinfo{author}{\bibfnamefont{R.}~\bibnamefont{Foot}},
  \bibinfo{author}{\bibfnamefont{H.~N.} \bibnamefont{Long}}, \bibnamefont{and}
  \bibinfo{author}{\bibfnamefont{T.~A.} \bibnamefont{Tran}},
  \bibinfo{journal}{Phys. Rev.} \textbf{\bibinfo{volume}{D50}},
  \bibinfo{pages}{R34} (\bibinfo{year}{1994}), \eprint{hep-ph/9402243}.

\bibitem[{\citenamefont{Long}(1996)}]{Hoang:1995vq}
\bibinfo{author}{\bibfnamefont{H.~N.} \bibnamefont{Long}},
  \bibinfo{journal}{Phys. Rev.} \textbf{\bibinfo{volume}{D53}},
  \bibinfo{pages}{437} (\bibinfo{year}{1996}), \eprint{hep-ph/9504274}.

\bibitem[{\citenamefont{Pleitez}(1996)}]{Pleitez:1994pu}
\bibinfo{author}{\bibfnamefont{V.}~\bibnamefont{Pleitez}},
  \bibinfo{journal}{Phys. Rev.} \textbf{\bibinfo{volume}{D53}},
  \bibinfo{pages}{514} (\bibinfo{year}{1996}), \eprint{hep-ph/9412304}.

\bibitem[{\citenamefont{Long}(1998)}]{Hoang:1997su}
\bibinfo{author}{\bibfnamefont{H.~N.} \bibnamefont{Long}},
  \bibinfo{journal}{Mod. Phys. Lett.} \textbf{\bibinfo{volume}{A13}},
  \bibinfo{pages}{1865} (\bibinfo{year}{1998}), \eprint{hep-ph/9711204}.

\bibitem[{\citenamefont{Dong et~al.}(2013)\citenamefont{Dong, Nguyen, and
  Soa}}]{Dong:2013ioa}
\bibinfo{author}{\bibfnamefont{P.~V.} \bibnamefont{Dong}},
  \bibinfo{author}{\bibfnamefont{T.~P.} \bibnamefont{Nguyen}},
  \bibnamefont{and} \bibinfo{author}{\bibfnamefont{D.~V.} \bibnamefont{Soa}},
  \bibinfo{journal}{Phys. Rev.} \textbf{\bibinfo{volume}{D88}},
  \bibinfo{pages}{095014} (\bibinfo{year}{2013}), \eprint{1308.4097}.

\bibitem[{\citenamefont{Pisano and Pleitez}(1992)}]{Pisano:1991ee}
\bibinfo{author}{\bibfnamefont{F.}~\bibnamefont{Pisano}} \bibnamefont{and}
  \bibinfo{author}{\bibfnamefont{V.}~\bibnamefont{Pleitez}},
  \bibinfo{journal}{Phys. Rev.} \textbf{\bibinfo{volume}{D46}},
  \bibinfo{pages}{410} (\bibinfo{year}{1992}), \eprint{hep-ph/9206242}.

\bibitem[{\citenamefont{Frampton}(1992)}]{PhysRevLett.69.2889}
\bibinfo{author}{\bibfnamefont{P.~H.} \bibnamefont{Frampton}},
  \bibinfo{journal}{Phys. Rev. Lett.} \textbf{\bibinfo{volume}{69}},
  \bibinfo{pages}{2889} (\bibinfo{year}{1992}).

\bibitem[{\citenamefont{Foot et~al.}(1993)\citenamefont{Foot, Hernandez,
  Pisano, and Pleitez}}]{Foot:1992rh}
\bibinfo{author}{\bibfnamefont{R.}~\bibnamefont{Foot}},
  \bibinfo{author}{\bibfnamefont{O.~F.} \bibnamefont{Hernandez}},
  \bibinfo{author}{\bibfnamefont{F.}~\bibnamefont{Pisano}}, \bibnamefont{and}
  \bibinfo{author}{\bibfnamefont{V.}~\bibnamefont{Pleitez}},
  \bibinfo{journal}{Phys. Rev.} \textbf{\bibinfo{volume}{D47}},
  \bibinfo{pages}{4158} (\bibinfo{year}{1993}), \eprint{hep-ph/9207264}.

\bibitem[{\citenamefont{Tonasse}(1996)}]{Tonasse:1996cx}
\bibinfo{author}{\bibfnamefont{M.~D.} \bibnamefont{Tonasse}},
  \bibinfo{journal}{Phys. Lett.} \textbf{\bibinfo{volume}{B381}},
  \bibinfo{pages}{191} (\bibinfo{year}{1996}), \eprint{hep-ph/9605230}.

\bibitem[{\citenamefont{Nguyen et~al.}(2000)\citenamefont{Nguyen, Ky, and
  Long}}]{Nguyen:1998ui}
\bibinfo{author}{\bibfnamefont{T.~A.} \bibnamefont{Nguyen}},
  \bibinfo{author}{\bibfnamefont{N.~A.} \bibnamefont{Ky}}, \bibnamefont{and}
  \bibinfo{author}{\bibfnamefont{H.~N.} \bibnamefont{Long}},
  \bibinfo{journal}{Int. J. Mod. Phys.} \textbf{\bibinfo{volume}{A15}},
  \bibinfo{pages}{283} (\bibinfo{year}{2000}), \eprint{hep-ph/9810273}.

\bibitem[{\citenamefont{Fonseca and Hirsch}(2016)}]{Fonseca:2016tbn}
\bibinfo{author}{\bibfnamefont{R.~M.} \bibnamefont{Fonseca}} \bibnamefont{and}
  \bibinfo{author}{\bibfnamefont{M.}~\bibnamefont{Hirsch}},
  \bibinfo{journal}{JHEP} \textbf{\bibinfo{volume}{08}}, \bibinfo{pages}{003}
  (\bibinfo{year}{2016}), \eprint{1606.01109}.

\bibitem[{\citenamefont{Georgi and Pais}(1979)}]{PhysRevD.19.2746}
\bibinfo{author}{\bibfnamefont{H.}~\bibnamefont{Georgi}} \bibnamefont{and}
  \bibinfo{author}{\bibfnamefont{A.}~\bibnamefont{Pais}},
  \bibinfo{journal}{Phys. Rev. D} \textbf{\bibinfo{volume}{19}},
  \bibinfo{pages}{2746} (\bibinfo{year}{1979}).

\bibitem[{\citenamefont{Schechter and Ueda}(1973)}]{PhysRevD.8.484}
\bibinfo{author}{\bibfnamefont{J.}~\bibnamefont{Schechter}} \bibnamefont{and}
  \bibinfo{author}{\bibfnamefont{Y.}~\bibnamefont{Ueda}},
  \bibinfo{journal}{Phys. Rev. D} \textbf{\bibinfo{volume}{8}},
  \bibinfo{pages}{484} (\bibinfo{year}{1973}).

\bibitem[{\citenamefont{Gupta and Mani}(1974)}]{PhysRevD.10.1310}
\bibinfo{author}{\bibfnamefont{V.}~\bibnamefont{Gupta}} \bibnamefont{and}
  \bibinfo{author}{\bibfnamefont{H.~S.} \bibnamefont{Mani}},
  \bibinfo{journal}{Phys. Rev. D} \textbf{\bibinfo{volume}{10}},
  \bibinfo{pages}{1310} (\bibinfo{year}{1974}).

\bibitem[{\citenamefont{Huang and Li}(1994)}]{Huang:1993qx}
\bibinfo{author}{\bibfnamefont{C.-S.} \bibnamefont{Huang}} \bibnamefont{and}
  \bibinfo{author}{\bibfnamefont{T.-J.} \bibnamefont{Li}},
  \bibinfo{journal}{Phys. Rev.} \textbf{\bibinfo{volume}{D50}},
  \bibinfo{pages}{2127} (\bibinfo{year}{1994}).

\bibitem[{\citenamefont{Huang and Li}(1995)}]{Huang:1994zg}
\bibinfo{author}{\bibfnamefont{C.-S.} \bibnamefont{Huang}} \bibnamefont{and}
  \bibinfo{author}{\bibfnamefont{T.-J.} \bibnamefont{Li}}, \bibinfo{journal}{Z.
  Phys.} \textbf{\bibinfo{volume}{C68}}, \bibinfo{pages}{319}
  (\bibinfo{year}{1995}).

\bibitem[{\citenamefont{Cao and Zhang}(2016)}]{Cao:2016uur}
\bibinfo{author}{\bibfnamefont{Q.-H.} \bibnamefont{Cao}} \bibnamefont{and}
  \bibinfo{author}{\bibfnamefont{D.-M.} \bibnamefont{Zhang}}
  (\bibinfo{year}{2016}), \eprint{1611.09337}.

\bibitem[{\citenamefont{Boucenna et~al.}(2015)\citenamefont{Boucenna, Valle,
  and Vicente}}]{Boucenna:2015zwa}
\bibinfo{author}{\bibfnamefont{S.~M.} \bibnamefont{Boucenna}},
  \bibinfo{author}{\bibfnamefont{J.~W.~F.} \bibnamefont{Valle}},
  \bibnamefont{and} \bibinfo{author}{\bibfnamefont{A.}~\bibnamefont{Vicente}},
  \bibinfo{journal}{Phys. Rev.} \textbf{\bibinfo{volume}{D92}},
  \bibinfo{pages}{053001} (\bibinfo{year}{2015}), \eprint{1502.07546}.

\bibitem[{\citenamefont{Huitu et~al.}(2019)\citenamefont{Huitu, Koivunen, and
  Kärkkäinen}}]{Huitu:2019mdr}
\bibinfo{author}{\bibfnamefont{K.}~\bibnamefont{Huitu}},
  \bibinfo{author}{\bibfnamefont{N.}~\bibnamefont{Koivunen}}, \bibnamefont{and}
  \bibinfo{author}{\bibfnamefont{T.~J.} \bibnamefont{Kärkkäinen}}
  (\bibinfo{year}{2019}), \eprint{1908.09384}.

\bibitem[{\citenamefont{Huitu and Koivunen}(2019)}]{Huitu:2019kbm}
\bibinfo{author}{\bibfnamefont{K.}~\bibnamefont{Huitu}} \bibnamefont{and}
  \bibinfo{author}{\bibfnamefont{N.}~\bibnamefont{Koivunen}},
  \bibinfo{journal}{JHEP} \textbf{\bibinfo{volume}{10}}, \bibinfo{pages}{065}
  (\bibinfo{year}{2019}), \eprint{1905.05278}.

\bibitem[{\citenamefont{Ponce et~al.}(2003)\citenamefont{Ponce, Giraldo, and
  Sanchez}}]{Ponce:2002sg}
\bibinfo{author}{\bibfnamefont{W.~A.} \bibnamefont{Ponce}},
  \bibinfo{author}{\bibfnamefont{Y.}~\bibnamefont{Giraldo}}, \bibnamefont{and}
  \bibinfo{author}{\bibfnamefont{L.~A.} \bibnamefont{Sanchez}},
  \bibinfo{journal}{Phys. Rev.} \textbf{\bibinfo{volume}{D67}},
  \bibinfo{pages}{075001} (\bibinfo{year}{2003}), \eprint{hep-ph/0210026}.

\bibitem[{\citenamefont{Dong et~al.}(2006)\citenamefont{Dong, Long, Nhung, and
  Soa}}]{Dong:2006mg}
\bibinfo{author}{\bibfnamefont{P.~V.} \bibnamefont{Dong}},
  \bibinfo{author}{\bibfnamefont{H.~N.} \bibnamefont{Long}},
  \bibinfo{author}{\bibfnamefont{D.~T.} \bibnamefont{Nhung}}, \bibnamefont{and}
  \bibinfo{author}{\bibfnamefont{D.~V.} \bibnamefont{Soa}},
  \bibinfo{journal}{Phys. Rev.} \textbf{\bibinfo{volume}{D73}},
  \bibinfo{pages}{035004} (\bibinfo{year}{2006}), \eprint{hep-ph/0601046}.

\bibitem[{\citenamefont{Deppisch et~al.}(2016)\citenamefont{Deppisch, Hati,
  Patra, Sarkar, and Valle}}]{Deppisch:2016jzl}
\bibinfo{author}{\bibfnamefont{F.~F.} \bibnamefont{Deppisch}},
  \bibinfo{author}{\bibfnamefont{C.}~\bibnamefont{Hati}},
  \bibinfo{author}{\bibfnamefont{S.}~\bibnamefont{Patra}},
  \bibinfo{author}{\bibfnamefont{U.}~\bibnamefont{Sarkar}}, \bibnamefont{and}
  \bibinfo{author}{\bibfnamefont{J.~W.~F.} \bibnamefont{Valle}},
  \bibinfo{journal}{Phys. Lett. B} \textbf{\bibinfo{volume}{762}},
  \bibinfo{pages}{432} (\bibinfo{year}{2016}), \eprint{1608.05334}.

\bibitem[{\citenamefont{Sen}(1985)}]{Sen:1983xj}
\bibinfo{author}{\bibfnamefont{A.}~\bibnamefont{Sen}}, \bibinfo{journal}{Phys.
  Rev. D} \textbf{\bibinfo{volume}{31}}, \bibinfo{pages}{900}
  (\bibinfo{year}{1985}).

\bibitem[{\citenamefont{Diaz et~al.}(2005)\citenamefont{Diaz, Martinez, and
  Ochoa}}]{Diaz:2004fs}
\bibinfo{author}{\bibfnamefont{R.~A.} \bibnamefont{Diaz}},
  \bibinfo{author}{\bibfnamefont{R.}~\bibnamefont{Martinez}}, \bibnamefont{and}
  \bibinfo{author}{\bibfnamefont{F.}~\bibnamefont{Ochoa}},
  \bibinfo{journal}{Phys. Rev.} \textbf{\bibinfo{volume}{D72}},
  \bibinfo{pages}{035018} (\bibinfo{year}{2005}), \eprint{hep-ph/0411263}.

\bibitem[{\citenamefont{Langacker}(1981)}]{Langacker:1980js}
\bibinfo{author}{\bibfnamefont{P.}~\bibnamefont{Langacker}},
  \bibinfo{journal}{Phys. Rept.} \textbf{\bibinfo{volume}{72}},
  \bibinfo{pages}{185} (\bibinfo{year}{1981}).

\bibitem[{\citenamefont{Sirunyan et~al.}(2019)}]{Sirunyan:2018nnz}
\bibinfo{author}{\bibfnamefont{A.~M.} \bibnamefont{Sirunyan}}
  \bibnamefont{et~al.} (\bibinfo{collaboration}{CMS}), \bibinfo{journal}{Phys.
  Lett.} \textbf{\bibinfo{volume}{B792}}, \bibinfo{pages}{345}
  (\bibinfo{year}{2019}), \eprint{1808.03684}.

\bibitem[{\citenamefont{Cárcamo~Hernández and
  King}(2019)}]{CarcamoHernandez:2019eme}
\bibinfo{author}{\bibfnamefont{A.~E.} \bibnamefont{Cárcamo~Hernández}}
  \bibnamefont{and} \bibinfo{author}{\bibfnamefont{S.~F.} \bibnamefont{King}}
  (\bibinfo{year}{2019}), \eprint{1903.02565}.

\bibitem[{\citenamefont{Gavela et~al.}(2009)\citenamefont{Gavela, Hambye,
  Hernandez, and Hernandez}}]{Gavela:2009cd}
\bibinfo{author}{\bibfnamefont{M.~B.} \bibnamefont{Gavela}},
  \bibinfo{author}{\bibfnamefont{T.}~\bibnamefont{Hambye}},
  \bibinfo{author}{\bibfnamefont{D.}~\bibnamefont{Hernandez}},
  \bibnamefont{and}
  \bibinfo{author}{\bibfnamefont{P.}~\bibnamefont{Hernandez}},
  \bibinfo{journal}{JHEP} \textbf{\bibinfo{volume}{09}}, \bibinfo{pages}{038}
  (\bibinfo{year}{2009}), \eprint{0906.1461}.

\bibitem[{\citenamefont{de~Salas et~al.}(2018)\citenamefont{de~Salas, Forero,
  Ternes, Tortola, and Valle}}]{deSalas:2017kay}
\bibinfo{author}{\bibfnamefont{P.~F.} \bibnamefont{de~Salas}},
  \bibinfo{author}{\bibfnamefont{D.~V.} \bibnamefont{Forero}},
  \bibinfo{author}{\bibfnamefont{C.~A.} \bibnamefont{Ternes}},
  \bibinfo{author}{\bibfnamefont{M.}~\bibnamefont{Tortola}}, \bibnamefont{and}
  \bibinfo{author}{\bibfnamefont{J.~W.~F.} \bibnamefont{Valle}},
  \bibinfo{journal}{Phys. Lett.} \textbf{\bibinfo{volume}{B782}},
  \bibinfo{pages}{633} (\bibinfo{year}{2018}), \eprint{1708.01186}.

\bibitem[{\citenamefont{Esteban et~al.}(2019)\citenamefont{Esteban,
  Gonzalez-Garcia, Hernandez-Cabezudo, Maltoni, and Schwetz}}]{Esteban:2018azc}
\bibinfo{author}{\bibfnamefont{I.}~\bibnamefont{Esteban}},
  \bibinfo{author}{\bibfnamefont{M.~C.} \bibnamefont{Gonzalez-Garcia}},
  \bibinfo{author}{\bibfnamefont{A.}~\bibnamefont{Hernandez-Cabezudo}},
  \bibinfo{author}{\bibfnamefont{M.}~\bibnamefont{Maltoni}}, \bibnamefont{and}
  \bibinfo{author}{\bibfnamefont{T.}~\bibnamefont{Schwetz}},
  \bibinfo{journal}{JHEP} \textbf{\bibinfo{volume}{01}}, \bibinfo{pages}{106}
  (\bibinfo{year}{2019}), \eprint{1811.05487}.

\end{thebibliography}

\end{document}